\documentclass[11pt]{article}
\usepackage{epsf}
\usepackage{latexsym}

\newcommand{\ap}[3]{{\sl Ann.~Phys.} {\bf #1} (19#2) #3}

\newcommand{\np}[3]{{\sl Nucl. Phys.} {\bf #1} (19#2)~#3}
\newcommand{\pl}[3]{{\sl Phys. Lett.} {\bf #1} (19#2) #3}
\newcommand{\pr}[3]{{\sl Phys. Rev.} {\bf #1} (19#2) #3}

\newcommand{\zp}[3]{{\sl Z. Phys.} {\bf #1} (19#2) #3}
\newcommand{\vj}[4]{{\sl #1~}{\bf #2} (19#3) #4}

\def\be{\begin{equation}}
\def\ee{\end{equation}}
\def\bea{\begin{eqnarray}}
\def\eea{\end{eqnarray}}

\makeatletter 
\def\secteqno{\@addtoreset{equation}{section}%
\def\theequation{\thesection.\arabic{equation}}} 
 
\def\endsecteqno{\def{theequation\{\@ifundefined{chapter}%
{\arabic{equation}}{\thechapter.\arabic{equation}}}} 
\makeatother

\newcommand{\bfi}[1]{\begin{figure}[#1]}
\newcommand{\efi}{\end{figure}}
\newcommand{\bpi}[2]{\begin{picture}(#1,#2)}
\newcommand{\epi}{\end{picture}}

\newcommand{\nn}{\nonumber\\}

\def\lhs{{\it l.h.s.\/}} 
\def\rhs{{\it r.h.s.\/}} 

\def\eg{e.g.}

\newcommand{\g}{\gamma}
\newcommand{\prop}{\Delta}
\newcommand{\propm}{{\Delta}_m}
\newcommand{\dsl}{{\not \! \partial}}
\renewcommand{\d}{\partial}
\newcommand{\T}{{\mathrm T}}
\newcommand{\Q}{{\mathrm Q}}
\newcommand{\B}{{\mathrm B}}
\newcommand{\A}{{\mathrm A}}
\newcommand{\I}{{\mathrm I}}
\newcommand{\F}{{\mathrm F}}
\newcommand{\OO}{{\mathcal O}}

\begin{document}

\pagestyle{empty} 
{\hfill \parbox{6cm}{\begin{center} 
                     MIT-CTP-2705\\ 
                     UG-FT-86/98  \\ 
                     hep-ph/9806451 \\
                     June 1998                   
                     \end{center}}} 
	      
\vspace*{1cm}                               
\begin{center} 
\large{\bf Techniques for one-loop calculations \\ }
\large{\bf in constrained differential renormalization} 
\vskip .6truein 
\centerline {\large 
      F. del \'Aguila, $^{a,}$\footnote{e-mail: faguila@goliat.ugr.es}
      A. Culatti, $^{a,b,}$\footnote{e-mail:  culatti@mvxpd5.pd.infn.it}
      R. Mu\~noz Tapia, $^{a,}$\footnote{e-mail: rmt@ugr.es} and
      M. P\'erez-Victoria $^{a,c,}$\footnote{e-mail: mpv@ugr.es}}  
\end{center} 
\vspace{.3cm} 
\begin{center}
{$^{a}$ 
Dpto. de F\'{\i}sica Te\'orica y del Cosmos,  
 Universidad de Granada, \\
 18071 Granada, Spain 
  \\} 
\vspace{.15cm} 
{$^{b}$  Dip. di Fisica, 
 Universit\`a di Padova, 35131 Padova, Italy \\} 
\vspace{.15cm} 
{$^{c}$  Center for Theoretical Physics, Laboratory for Nuclear
Science and \\ Department of Physics, 
Massachusetts Institute of
Technology,  \\ Cambridge, MA 02139, USA} 
\end{center}
\vspace{1.5cm} 
 
\centerline{\bf Abstract} 
\medskip 

We describe in detail the constrained procedure
of differential renormalization and
develop the techniques required for one-loop calculations.
As an illustration
we renormalize Scalar QED and show that the two-, three- 
and four-point Ward identities are automatically satisfied.

\vspace*{1cm}

\pagestyle{plain}
\setcounter{footnote}{0}
\secteqno

\section{Introduction}

Differential renormalization~(DR)~\cite{FJL} is a method of 
renormalization in coordinate space that yields directly finite 
Green functions without intermediate regularization or explicit
counterterms. It has proved to be quite simple and powerful in a 
number of applications~\cite{DRapplications} (see also
Refs.~\cite{DRdevelopments,DRcounterterms,massiveDR} for formal 
developments). 
Standard DR manipulates singular objects as if they were
well-defined, expresses them in terms of simple singular functions,
and substitutes these by their renormalized value.
The renormalized functions contain arbitrary dimensionful constants 
which play the role of renormalization scales
and carry all the ambiguity inherent to the formal 
manipulation of singular expressions. 
In symmetric theories, these  
scales can be adjusted in such a way that the Ward identities for 
renormalized Green functions are satisfied. For practical purposes, 
however, one would prefer a procedure that directly 
rendered symmetric expressions,
without having to impose the Ward identities after each calculation.

Such a procedure, {\em constrained differential renormalization} 
(CDR), was introduced in Ref.~\cite{CDR} at the one-loop 
level. The idea is to impose that the renormalized expressions be 
compatible with a minimal set of consistent formal 
manipulations ({\em rules}). As a result, the ambiguities and 
arbitrary renormalization scales of DR are fixed and 
the resulting renormalized
Green functions automatically preserve Ward identities. This was
explicitly shown in Ref.~\cite{CDR} for a variety of cases in 
abelian gauge theories. Supersymmetry was also preserved in the
calculation of supergravity corrections to the anomalous
magnetic moment of a charged lepton~\cite{g2,SUSY}.
The method also works for non-abelian gauge theories, as shown 
in Ref.~\cite{QCD}.
The extension of the constrained method to higher orders is more 
involved and needs further study. At any rate, it requires 
a good understanding and systematization of the one-loop order.

The purpose of this paper is to describe CDR in detail and
introduce the techniques required for one-loop calculations in any 
renormalizable theory in four dimensions. As an example, we renormalize
(to one loop) Scalar QED, which contains all possible kinds 
of diagrams and derivative structures that appear in 
renormalizable theories.
The plan of the paper is the following. Section~2 contains 
the procedure and rules of CDR. 
We first motivate them taking
as a guiding example the vertex Ward identity in QED. Then we 
discuss how to expand any one-loop Feynman 
graph in a complete set of {\em basic functions\/} and we give 
the renormalization rules. Finally, the rules are used to determine 
the renormalization of the singular basic functions. 
The reader interested
in practical calculations only needs the tables in this
section. The full derivation
is given in Appendix A. 
In Section~3 we apply the method to Scalar QED. We 
renormalize the two-, three- and four-point 1PI Green functions
and verify that they satisfy the Ward identities.  
Section 4 is devoted to conclusions. In Appendix B we collect
the Fourier transforms of CDR expressions. 
An efficient computer code performing all
operations automatically (in momentum space) 
is available~\cite{program}.

\section{Constrained differential renormalization}

In general, the bare expressions of Feynman diagrams in
coordinate space are too singular
at coincident points to behave as tempered distributions. The role
of renormalization is to consistently 
replace these ``singular'' expressions 
by ``regular'' ones.
To carry out this program,
we first reduce each Feynman graph
to a sum of (singular) basic funtions and their derivatives, and
then define the renormalized diagram as the corresponding sum of the
renormalized basic functions. CDR is essentially a
set of rules for manipulating the singular expressions and
fixing the renormalization of a complete set of basic functions.
These rules are chosen in such a way that Ward identities are
respected. We specify the rules below, after motivating them with a
simple example. Then, we define the (one-loop) basic
functions and apply the rules to renormalize them.

\subsection{Motivation of the rules}

Ward identities among Green functions are derived from the symmetries 
of the action using general properties of Quantum Field Theory, as
gathered in Schwinger's action principle~\cite{Schwinger}. 
These properties are only formal for bare Green functions because 
they involve ill-defined expressions. Regularization and 
renormalization can invalidate them, breaking the symmetries at the 
quantum level. In absence of anomalies, the symmetry can 
be restored adding adequate finite local 
counterterms, as a consequence of the quantum action 
principle~\cite{Lowenstein,Piguet}. However, in real calculations
it is preferable to have an invariant procedure rendering
directly symmetric expressions. 
This is the case of dimensional regularization 
with minimal substraction in gauge 
theories~\cite{dimensional,Breitenlohner}. 
In perturbation theory, the fulfilment of the   
Ward identities (and of the action principle, in general)
relies heavily on the following 
property~\cite{Breitenlohner,diagrammar}:
the application of the kinetic differential operator
to the propagator corresponding to some line in a Feynman
graph is equivalent to the contraction of the line to
a point. In other words, the fact that the free propagators 
are Green functions of the corresponding quadratic lagrangian
must not be spoiled by the renormalization procedure.
Actually, the corresponding proof for gauge theories 
in Ref. \cite{Breitenlohner} is  based on the
fact that this property holds for the
dimensionally regulated  diagrams.
We shall see that the propagator equations also play a central
role in CDR. Although no explicit regulator is available
in DR, one can impose simple conditions on the renormalized
expressions that guarantee the validity of the propagator equations 
and of other formal identities. To see what kind of 
constraints one must impose, let us consider the case of the 
vertex Ward identity in massless QED,
\begin{equation}
(\d_\mu^x - \d_\mu^y) V_\mu(x,y) = 
i e [ \Sigma(x) \delta(y) - \Sigma(y) \delta(x) ]\, , 
\label{WI}
\end{equation}
where $V_\mu$ is the 1PI photon-electron-electron vertex, $\Sigma$ 
is the electron self-energy and $x \equiv x_1 - x_2,~ 
y \equiv x_2 - x_3$, 
with $x_{1,2,3}$ the external points ($x_2$ being the point 
attached to the photon)\footnote{For convenience when
treating the general case, we use in this paper
the variable $y$ with a minus sign with respect to 
the one in Refs.~\cite{CDR,g2,SUSY,polonia}.}. 
For bare functions this Ward identity is easy to derive formally 
at one loop. Using the Feynman rules in Ref.~\cite{polonia} 
(we work in euclidean space and in the Feynman gauge),  
\begin{equation}
\begin{array}{lll}
(\d_\mu^x - \d_\mu^y) V_\mu(x,y) & = &
(\d_\mu^x - \d_\mu^y) [ (-ie)^3 \g_\alpha \dsl^x \prop(x) \g_\mu
   \dsl^y \prop(y) \g_\alpha \prop(x+y) ] \\ 
& = &
ie^3 [  \g_\alpha \dsl^x \dsl^x \prop(x) 
   \dsl^y \prop(y) \g_\alpha \prop(x+y) \\ 
& & - \g_\alpha \dsl^x \prop(x) 
   \dsl^y \dsl^y \prop(y) \g_\alpha \prop(x+y) \\ 
& & + \g_\alpha \dsl^x \prop(x) \g_\mu 
   \dsl^y \prop(y) \g_\alpha (\d_\mu^x - \d_\mu^y) \prop(x+y) ] \\ 
& = &
ie^3 [ - \delta(x) \g_\alpha \dsl^y \prop(y) \g_\alpha \prop(x+y) \\  
& & + \delta(y) \g_\alpha \dsl^x \prop(x) \g_\alpha \prop(x+y) ] \\ 
& = & 
i e [ - \delta(x) \Sigma(y) + \delta(y) \Sigma(x) ]\, , \\ 
\end{array}
\label{VWI}
\end{equation}
where $\prop(x)=\frac{1}{4\pi^2} \frac{1}{x^2}$ 
is the massless Feynman propagator,  
$\d_\mu^x$ stands for $\frac{\d}{\d x_\mu}$ and
$\Box=\d_\mu \d_\mu$.
The manipulations used in  Eq.~(\ref{VWI}) are 
only formal because we are dealing with singular expressions 
(at $x = y$). 
We have first performed the linear change of variables described
above, and used the Leibnitz rule to make the 
external derivatives act on individual propagators. 
Then we have used the propagator equation for 
a massless fermion (which is equivalent
to replacing $\dsl \dsl$ by $\Box$ and using the
scalar propagator equation, $\Box \prop = -\delta$).  
The symmetry of $\prop(x+y)$ under the interchange of $x$ and $y$
makes $(\d_\mu^x - \d_\mu^y) \prop(x+y)$ vanish; finally, the 
definition of the Dirac delta function justifies the last 
equality.
When these ill-defined expressions are regularized or 
replaced by finite ones (renormalized), some of these manipulations 
may not be justified. For example, if the fermion propagator 
were na\"\i vely regulated {\`a} la Pauli-Villars, 
$S^{PV}(x) = \dsl^x (\prop (x) - \prop _\Lambda (x))$, 
with $\prop _\Lambda$ the Feynman propagator of mass $\Lambda$,
the propagator equation would be modified: 
$\dsl S^{PV} = - \Lambda ^2 \prop _\Lambda (x)$, 
and the third equality in Eq.~(\ref{VWI}) would no longer hold. 
Other two examples studied in the DR literature are 
the regulators introduced in
Refs.~\cite{DRcounterterms} and~\cite{FJL}, respectively. 
In the former, the modification 
of the propagator invalidates the propagator equation; 
in the latter the short distance expansion invalidates the 
change of variables and the direct use of the delta 
function\footnote{
Both methods were adequate to study the relation between DR 
and the usual counterterm approach, although
the spirit of DR is precisely to avoid the use of such regulators.}. 
Our aim is to find a DR scheme automatically 
preserving the Ward identities 
for renormalized amplitudes. One may require the whole 
Eq.~(\ref{WI}) to be satisfied by the corresponding 
renormalized expressions, as we essentially did in Ref.~\cite{g2}
to relate diagrams with different topology. However this has to be 
improved because diagrams are expected to be renormalized 
independently of the particular combination in which they 
appear in the Ward identities. Instead, we
demand that the manipulations performed to derive 
Eq.~(\ref{VWI}) be always valid
for renormalized expressions. In other words, we require
that DR commutes with this kind of manipulations. 
In what follows we show how this can be 
done consistently at one loop. Indeed, a few rules 
will be sufficient to fix the ambiguities of standard DR and
to ensure the validity of the Ward identities at the same time.

\subsection{Rules for constrained differential renormalization}

We distinguish two kinds of rules. The first ones simply state that
renormalization commutes with
algebraic identities that allow to perform
``straightforward'' manipulations, like sums of terms, Dirac
algebra (in four dimensions), or application of the
Leibnitz rule. 
With these manipulations one can express any
one-loop Feynman diagram in terms of a set of basic 
functions that will be defined in the next subsection.
The rules of the second kind 
extend well-defined
identities of distribution theory to more
singular expressions. 
They will be used to determine the renormalization of 
the basic functions.
Let us enumerate them (the first two rules are 
essentially the prescriptions of the method of DR)~\cite{CDR}:
\begin{enumerate}
  \item {\em Differential reduction}: 
  singular expressions are substituted by derivatives
  of regular ones. 
  \label{R1} This can be done in two steps:
     \begin{enumerate}
       \item Functions with singular behaviour worse than 
        logarithmic ($ \sim x^{-4}$) are reduced to derivatives 
        of logarithmically
        singular functions without introducing extra dimensionful
        constants. \label{R1a}
       \item Logarithmically singular functions are written as
        derivatives of regular functions. The
        usual DR identity~\cite{FJL}
\be
  \left[\frac{1}{x^4} \right]^R
  = -\frac{1}{4}  \Box \frac{\log x^2 M^2}{x^2}\, ,
\label{DRidentity}
\ee 
        is sufficient to one loop. 
        This identity also applies to massive
        expressions when they are expanded
        in the mass.
        Eq.~(\ref{DRidentity}) introduces the unique 
        dimensionful constant of
        the whole process, $M$, which has dimensions of
        mass and plays the role of the
        renormalization group scale. \label{R1b}
      \end{enumerate}   
  \item {\em Formal integration by parts}: derivatives act
   formally by parts on test functions. \label{R2} In particular,
\be
  [\d F]^R = \d F^R \, ,
\ee
   where $F$ is an arbitrary function and $R$ stands
   for renormalized.
  \item {\em Delta function renormalization rule}: \label{R3}
\be
  [F(x,x_1,...,x_n) \delta(x-y)]^R =  [F(x,x_1,...,x_n)]^R
  \delta(x-y)\, .
\ee 
  \item The general validity of the {\em propagator equation}:
   \label{R4}
\be 
  \left[ F(x,x_1,...,x_n) (\Box^x - m^2) \propm(x) \right]^R = 
  \left[ F(x,x_1,...,x_n) (- \delta(x)) \right]^R\, ,
  \label{masspropeq}
\ee
   where $\propm(x) = \frac{1}{4\pi^2} \frac{m K_1(mx)}{x}$ and
   $K_1$ is a modified Bessel function~\cite{Abramowitz}.
\end{enumerate}
Rule~\ref{R1}, combined with rule~\ref{R2},  
reduces the ``degree of singularity'',
connecting singular and regular expressions. 
Forbidding the introduction
of dimensionful scales outside logarithms, we
completely fix the scheme. 
The last three rules are valid mathematical identities among 
tempered distributions when applied to a well-behaved enough
function $F$. The rules formally extend their range of 
applicability to arbitrary functions. They require that 
these identities commute with the process of renormalization.
Rule~\ref{R2} is essential to make sense of rule~\ref{R1}, for 
otherwise the right-hand-side of it would not be a well-defined
distribution. 
Rule~\ref{R4} connects functions
with different number of propagators and has actually been applied
in the literature to reduce three-point functions to
two-point functions, thus allowing the use of the DR 
identity~(\ref{DRidentity}). As we shall see 
in the next subsection,
it has further important implications. Note that the massive
propagator equation~(\ref{masspropeq}) carries the necessary 
information 
for the propagators appearing in usual  
theories. Indeed, the
corresponding equation for a massless scalar propagator
is just the limit $m \rightarrow 0$ of Eq.~(\ref{masspropeq}),
the equation for a fermionic propagator follows from it, 
after the use of $\dsl \dsl = \Box$, and
the equation for the propagator of a gauge boson in a 
general covariant gauge can be derived from the first two terms of 
the mass expansion of Eq.~(\ref{masspropeq}) (see Ref.~\cite{CDR}).

All these rules allow to renormalize any one-loop Feynman graph. 
Other possible manipulations can be incompatible with them and  
introduce ambiguities.
One can still perform them as long 
as one keeps track of the arbitrary
local terms they introduce. These local terms can then be fixed 
imposing consistency with the renormalization rules. In particular, 
two dangerous
operations are explicit differentiation (which is not well-defined
at coincident points)\footnote{
Taking derivatives is essential to obtain the DR identities, but the
corresponding ambiguities are eventually fixed by the 
CDR rules.} and index contraction~\cite{CDR,polonia}, as we shall
see explicitly later on.
These two manipulations do not commute in general with CDR.

The actual procedure of renormalization involves
two steps:  (1)~express a Feynman diagram in terms of basic 
functions and  (2)~replace the basic functions by their
renormalized value.
The renormalization of the basic functions is done once and for all 
and does not depend on the particular calculation. 
This is the subject of the
next subsection. Let us now comment on the first step.
At one loop, a Feynman graph in coordinate space contains
products of propagators with differential operators acting on some 
of them and, possibly, delta functions. 
It can also include constant objects like group factors, 
gamma matrices, the metric and the Levi-Civita tensor. 
To evaluate the diagram, we first make 
a convenient change of variables (see below). 
Then, we perform all the (Dirac) algebra. In particular,
we do all possible contractions. This is a necessary
prescription because index contraction does not commute
with CDR. This prescription leads to universal renormalized 
basic functions.
Finally, the Leibnitz rule is used to reorder
the derivatives in each term, so that they act either on the whole 
product or on the last propagator. Symmetries among the 
space-time points can also
be safely used as, for example, in 
$F(x,y) \d_\mu^y \prop(x+y) = F(x,y) \d_\mu^x \prop(x+y)$.

\subsection{Renormalization of basic functions}

We shall assume that a Feynman-like gauge is chosen 
for the gauge fields,
so that their propagators are proportional to the scalar 
Feynman propagator. The formalism can be directly extended to 
general covariant gauges, as was indicated in Ref.~\cite{CDR}.
Up to delta functions, any one-loop 1PI graph 
is a linear combination of products of scalar propagators 
$\prop_{m_i}(x_i-x_{i+1})$,
with differential operators $\OO^{x_i}$ acting on them. 
All points $x_i$ are external and appear in a cyclic way.
Therefore, using the Leibnitz rule, each term of 
the graph can be written 
(ignoring constants and delta functions) as a sum of
total derivatives of the whole product of propagators, with all 
the internal derivatives acting on only one adequately chosen
propagator:
\bea
 \lefteqn{\OO_1^{x_1} \prop_{m_1}(x_1-x_2) \OO_2^{x_2} 
 \prop_{m_2}(x_2-x_3) \cdots \OO_n^{x_n} \prop_{m_n}(x_n-x_1) } \nn 
  & = & 
  \OO_1^{z_1} \prop_{m_1}(z_1) \OO_2^{z_2} \prop_{m_2}(z_2) \cdots
  \OO_n^{-z_1} \prop_{m_n}(z_1+z_2+ \cdots +z_{n-1}) \nn 
  & = & 
  \sum_i \OO_{ext(i)} [ \prop_{m_1}(z_1) \prop_{m_2}(z_2) \cdots 
  \OO_{int(i)}^{z_1} \prop_{m_n}(z_1+z_2+ \cdots +z_{n-1})]\, .
\label{bfreduction}
\eea
In the first line we have
performed a convenient change of 
variables: $z_1=x_1-x_2$, $z_2=x_2-x_3$, \dots, $z_{n-1}=x_{n-1}-x_n$, 
$z_n=x_n$. This eliminates one variable ($z_n$) due to translational
invariance. Besides, since every $z_i$ appears only in one of the first
$n-1$ propagators and in the last one, it is straightforward to use
the Leibnitz rule to make all derivatives act on the last propagator
and obtain the second equality.
Because of rule~\ref{R2}, the renormalization of the
graph reduces to renormalizing expressions of the form
\bea
  \lefteqn{\F^{(n)}_{m_1 m_2 \dots m_{n-1} m_n}[\OO](z_1,z_2, 
  \dots, z_{n-1}) \equiv} \nn 
  & & \makebox[0.5cm]{}\prop_{m_1}(z_1) \prop_{m_2}(z_2)
  \cdots \prop_{m_{n-1}}(z_{n-1})
  \OO^{z_1} \prop_{m_n}(z_1 + z_2 + \cdots +z_{n-1})\, , 
\eea
which we call {\em basic functions}. For massless basic functions 
we shall suppress the mass subindices. 
In renormalizable theories,
singular one-loop diagrams involve basic functions with at most
four propagators and no more derivatives than propagators. All
the singular basic functions for this class of theories 
(in the Feynman gauge) are displayed in Table~1. 
We denote $\A$, $\B$, $\T$ and $\Q$ the basic
functions with one, two, three and four propagators, respectively.  
In the following we shall use $x$, $y$, $z$ to denote $z_{1,2,3}$, 
respectively, and
assume that they are the (ordered) arguments of the basic functions
unless otherwise specified (\eg, $\T[\Box] \equiv \T[\Box](x,y)$).
\begin{table}
\begin{flushleft}
  \begin{displaymath}
  \begin{array}{l|c|c|c|c}
    & \makebox[3cm]{logarithmic} & \makebox[3cm]{linear} 
      & \makebox[3cm]{quadratic} & \makebox[3cm]{cubic} \\
    \hline 
    \makebox[2cm][l]{1 prop.} & & & \A_m[1] & \A_m[\d_\mu] \\
    \hline 
    \makebox[2cm][l]{2 props.} & \B_{m_1m_2}[1] & \B_{m_1m_2}[\d_\mu] 
      & \B_{m_1m_2}[\Box] & \\
      & & & \B_{m_1m_2}[\d_\mu\d_\nu] & \\
    \hline 
    \makebox[2cm][l]{3 props.} & \T_{m_1m_2m_3}[\Box]  & 
      \T_{m_1m_2m_3}[\Box\d_\mu] & & \\
      & \T_{m_1m_2m_3}[\d_\mu\d_\nu] & 
      \T_{m_1m_2m_3}[\d_\mu\d_\nu\d_\rho] & & \\
    \hline 
    & \Q_{m_1m_2m_3m_4}[\Box\Box] & & & \\
    \makebox[2cm][l]{4 props.} & \Q_{m_1m_2m_3m_4}[\Box \d_\mu\d_\nu] 
    & & & \\
    & \Q_{m_1m_2m_3m_4}[\d_\mu\d_\nu\d_\rho\d_\sigma] & & & \\
    \hline
  \end{array}
  \end{displaymath}
\end{flushleft}
\caption{Singular basic functions for renormalizable theories in
four dimensions. Lines are ordered according to the number of
propagators and columns according to the degree of 
singularity. The function $\A$, that appears in
tadpoles, is defined as $\A_m[\OO]=\OO \prop_m(x) \delta(x)$.}
\end{table}
The renormalization of these functions is carried out using the 
rules of CDR \ref{R1}-\ref{R4}. Renormalized one- and
two-point functions are gathered in Tables~2 and~3, and three-
and four-point functions, in Table~4. For the reader's convenience
the massless A and B functions are given in Table~2, although
they can be obtained from the massive ones in Table~3, taking
the appropriate limit.
The expressions for T and Q functions in Table~4 apply
directly for any value of the masses.
The Fourier transforms are collected in Appendix~B. 
The reader interested in applications only needs
the expressions in Tables~2, 3 and~4 for coordinate
space or Tables~5, 6 and~7 in Appendix~B 
for momentum space calculations. 
\begin{table}[h]
\begin{center}
\begin{math}
\begin{array}{l}
\hline \\
  \A^R[1]  =  0  \\ \\
  \A^R[\d_\mu]  =  0 \\ \\
  \B^R[1]  =  - \frac{1}{64\pi^4} 
    \Box \frac{\log x^2 M^2}{x^2}  \\ \\
  \B^R[\d_\mu]  =  \frac{1}{2} \d_\mu \B^R[1]  \\ \\
  \B^R[\Box]  =  0  \\ \\
  \B^R[\d_\mu\d_\nu]  =   \frac{1}{3} 
    (\d_\mu\d_\nu - \frac{1}{4}\delta_{\mu\nu} \Box) 
    \B^R[1] +
    \frac{1}{288\pi^2} (\d_\mu\d_\nu - \delta_{\mu\nu} \Box)
    \delta(x) \\ \\
\hline \\
\end{array} 
\end{math}
\end{center}
\caption{Renormalized expressions of massless
one- and two-point basic functions.}
\end{table}

%
\begin{table}[h]
\begin{center}
\begin{math}
\begin{array}{l}
\hline \\
  \A_m^R[1]  =  \frac{1}{16\pi^2} m^2 
    (1-\log \frac{\bar{M}^2}{m^2}) \delta(x)  \\ \\
  \A_m^R[\d_\mu]  =  0 \\ \\
  \B^R_{m_1m_2}[1]  =   
    \frac{1}{32\pi^4} \left\{ \frac{m_1m_2}{m_1+m_2}
    [\Box-(m_1+m_2)^2]
    \frac{K_0(m_1 x) K_1(m_2 x) + K_0(m_2 x) K_1(m_1 x)}{x} 
    \right. \\
    \left.  \makebox[1cm]{} + 
    2 \pi^2 \left(\log \frac{\bar{M}^2}{m_1m_2}
    + \frac{m_1-m_2}{m_1+m_2} \log \frac{m_2}{m_1}\right)
    \delta(x)  \right\} \\ \\
  \B_{m_1m_2}^R[\d_\mu]  = 
    \frac{1}{2} \d_\mu \B_{m_1m_2}^R[1]
    + \frac{1}{64\pi^4} \left[
    m_2^2 K_0(m_2 x) \d_\mu \frac{m_1 K_1(m_1 x)}{x}
    -  m_1^2 K_0(m_1 x) \d_\mu \frac{m_2 K_1(m_2 x)}{x} 
    \right] \\ \\
  \B_{m_1m_2}^R[\Box]  = 
    m_2^2 \B_{m_1m_2}^R[1] - A_{m_1}^R[1] \\ \\
  \B_{m_1m_2}^R[\d_\mu\d_\nu]  = 
    \frac{1}{2} \left( \d_\mu \B_{m_1m_2}^R[\d_\nu] -
    \d_\nu \B_{m_2m_1}^R[\d_\mu] \right) +
    \frac{1}{8} \delta_{\mu\nu} \left( \B_{m_1m_2}^R[\Box] +
    \B_{m_2m_1}^R[\Box] \right) \\    
    \makebox[1cm]{} +
    \frac{1}{3} (\d_\mu\d_\nu-\frac{1}{4}\delta_{\mu\nu}\Box)
    \B_{m_1m_2}^R[1] \\
    \makebox[1cm]{} +
    \frac{1}{192\pi^4} \left\{ \left[
    m_1^2 \frac{m_1 K_1(m_1 x)}{x} 
    (\d_\mu\d_\nu-\frac{1}{4}\delta_{\mu\nu}\Box) K_0(m_2 x) 
    \right. \right. \\
    \left. \makebox[1cm]{} + m_2^2 \frac{m_2 K_1(m_2 x)}{x} 
    (\d_\mu\d_\nu-\frac{1}{4}\delta_{\mu\nu}\Box) K_0(m_1 x)
    \right] \\
    \makebox[1cm]{} - \left[
    m_1^2 K_0(m_1 x) (\d_\mu\d_\nu-\frac{1}{4}\delta_{\mu\nu}\Box)
    \frac{m_2 K_1(m_2 x)}{x} \right. \\
    \left. \left. \makebox[1cm]{} + 
    m_2^2 K_0(m_2 x) (\d_\mu\d_\nu-\frac{1}{4}\delta_{\mu\nu}\Box)
    \frac{m_1 K_1(m_1 x)}{x} \right] \right\} \\
    \makebox[1cm]{} +
    \frac{1}{16\pi^2} \left[ \frac{1}{18} 
    (\d_\mu\d_\nu-\delta_{\mu\nu}\Box) +  
    \frac{1}{8} (m_1^2+m_2^2) \delta_{\mu\nu} \right] \delta(x)
     \\ 
    \mbox{} \\
\hline    
\end{array} 
\end{math}
\end{center}
\caption{Renormalized expressions of massive
one- and two-point basic functions.}
\end{table}
\begin{table}[h]
\begin{center}
\begin{math}
\begin{array}{l}
\hline \\
  \T_{m_1m_2m_3}^R[\Box]  = 
    m_3^2 \T_{m_1m_2m_3}[1] - \B_{m_1m_2}^R[1] \delta(x+y) \\ \\
  \T_{m_1m_2m_3}^R[\d_\mu\d_\nu]  =  
    \T_{m_1m_2m_3}[\d_\mu\d_\nu-\frac{1}{4}\delta_{\mu\nu}\Box]
    + \frac{1}{4} \delta_{\mu\nu} \T_{m_1m_2m_3}^R[\Box]
    -\frac{1}{128\pi^2} \delta_{\mu\nu} \delta(x)\delta(y) \\ \\
  \T_{m_1m_2m_3}^R[\Box\d_\mu]  =  
    m_3^2 \T_{m_1m_2m_3}[\d_\mu] - \B^R_{m_1m_2}[\d_\mu] 
    \delta(x+y) - \d_\mu^y \left( \B^R_{m_1m_2}[1] \delta(x+y)
    \right) \\ \\  
  \T_{m_1m_2m_3}^R[\d_\mu\d_\nu\d_\rho]  =  
    \T_{m_1m_2m_3}[\d_\mu\d_\nu\d_\rho-\frac{1}{6}
    (\delta_{\mu\nu} \d_\rho + \delta_{\mu\rho} \d_\nu
    +\delta_{\nu\rho} \d_\mu) \Box] \\
    \makebox[1cm]{}
    + \frac{1}{6} \left(
    \delta_{\mu\nu} \T_{m_1m_2m_3}^R[\Box\d_\rho] +
    \delta_{\mu\rho} \T_{m_1m_2m_3}^R[\Box\d_\nu] +
    \delta_{\nu\rho} \T_{m_1m_2m_3}^R[\Box\d_\mu] \right) \\
    \makebox[1cm]{}
    - \frac{1}{576\pi^2} \left(
    \delta_{\mu\nu} (\d_\rho^x+\d_\rho^y) +
    \delta_{\mu\rho} (\d_\nu^x+\d_\nu^y) +
    \delta_{\nu\rho} (\d_\mu^x+\d_\mu^y) \right)
    \left(\delta(x)\delta(y)\right) \\ \\
  \Q_{m_1m_2m_3m_4}^R[\Box\Box]  = 
    m_4^2 \Q_{m_1m_2m_3m_4}[\Box] - 
    \Box^z \left(\T_{m_1m_2m_3}[1] \delta(x+y+z) \right) \\
    \makebox[1cm]{}
    - 2 \d_\rho^z \left(
    \T_{m_1m_2m_3}[\d_\rho] \delta(x+y+z) \right) -
    \T_{m_1m_2m_3}^R[\Box] \delta(x+y+z) \\ \\
  \Q_{m_1m_2m_3m_4}^R[\Box\d_\mu\d_\nu]  = 
    m_4^2 \Q_{m_1m_2m_3m_4}[\d_\mu\d_\nu] -
    \d_\mu^z \d_\nu^z 
    \left(\T_{m_1m_2m_3}[1] \delta(x+y+z) \right) \\
    \makebox[1cm]{}  -
    \d_\mu^z \left(
    \T_{m_1m_2m_3}[\d_\nu] \delta(x+y+z) \right) -
    \d_\nu^z \left(
    \T_{m_1m_2m_3}[\d_\mu] \delta(x+y+z) \right) \\
    \makebox[1cm]{} -
    \T_{m_1m_2m_3}^R[\d_\mu\d_\nu] \delta(x+y+z) \\ \\
  \Q_{m_1m_2m_3m_4}^R[\d_\mu\d_\nu\d_\rho\d_\sigma]  =  
    \Q_{m_1m_2m_3m_4}[\d_\mu\d_\nu\d_\rho\d_\sigma -
    \frac{1}{24}(\delta_{\mu\nu}\delta_{\rho\sigma} +
    \delta_{\mu\rho}\delta_{\nu\sigma}+
    \delta_{\mu\sigma}\delta_{\nu\rho}) \Box\Box] \\
    \makebox[1cm]{} +
    \frac{1}{24} (\delta_{\mu\nu}\delta_{\rho\sigma} +
    \delta_{\mu\rho}\delta_{\nu\sigma}+
    \delta_{\mu\sigma}\delta_{\nu\rho})
    \left(\Q_{m_1m_2m_3m_4}^R[\Box\Box] + 
    \frac{5}{96\pi^2} \delta(x)\delta(y)\delta(z)
    \right) \\ \\
\hline
\end{array}
\end{math}
\end{center}
\caption{Renormalized expressions of three- and
four-point basic functions. The basic functions
with traceless differential operators are directly
finite.}
\end{table}
Note that it is important in our procedure to distinguish
basic functions with
contracted and uncontracted differential operators, because
contraction of Lorentz indices does not in general commute 
with CDR.
For instance, from Table~4~\cite{CDR,polonia},
\be
  \T^R[\Box]=\left[\delta_{\mu\nu} \T[\d_\mu \d_\nu]\right]^R
  \not = \delta_{\mu\nu} \T^R[\d_\mu\d_\nu] \, .
\label{traceCDR}
\ee
This justifies our prescription of contracting indices before 
identifying the basic functions.
In the following we briefly describe
how these tables have been obtained. The detailed derivation can be
found in Appendix~\ref{Appren}.

In general the renormalization of basic functions proceeds in
two steps. First, a differential equation is solved for non-singular
points, in order to express the singular basic functions 
as derivatives of well-behaved functions. 
A useful trick in the case of
complex tensor structures is to decompose them into
trace and traceless parts: the former carries the
leading singularity but is simpler, while the latter
is less singular.  
Second,
the arbitrary local terms are determined according to the CDR rules.
Rule~\ref{R1} is actually just an initial condition for
the local terms, while
the remaining rules relate the local terms of different basic
functions. Basic functions with simple
enough tensor structures
can be directly expressed in terms of other 
basic functions that have been previously renormalized.

Let us consider first the renormalization of $\B_{m_1 m_2}[1]$ and 
$\A_m[1]$. We start with the massless case. 
Direct application of rule~\ref{R1} gives
\be
  \B^R[1] = \frac{1}{(4\pi^2)^2} \left[\frac{1}{x^4}\right]^R =
  - \frac{1}{64\pi^4} \Box \frac{\log x^2 M^2}{x^2} \, ,
\label{BR1}
\ee
which is the standard DR identity.
The fact that $\A[1] = \prop(x) \delta(x)$ is local, together with 
power counting, implies that the most general renormalized value of
this function is of the form
\be
  \A^R[1]= (a \Box + \mu^2) \delta(x) \, .
\label{AR1}
\ee
Rule~\ref{R1a} tells us not to introduce the 
dimensionful constant $\mu$, 
so the second term in the equation above vanishes.
Eq.~(\ref{BR1}) and $\mu=0$ in Eq.~(\ref{AR1}), based on
rule~\ref{R1}, are the initial
conditions of the renormalization process. This rule is not needed 
anymore. As it is shown in Appendix~\ref{Appren}, 
rule~\ref{R3} implies
that the first term in Eq.~(\ref{AR1}) must also vanish, so
we have 
\be
  \A^R[1]=0\, .
\ee
For massive basic functions we use recurrence relations among 
modified Bessel functions to obtain the expressions for non-singular
points (see Ref.~\cite{massiveDR,Abramowitz} and Appendix~C of 
Ref.~\cite{g2}).
The local terms of $\B_{m_1m_2}^R[1]$ and $\A_m^R[1]$
are fixed so that in the massless
limit we recover 
$\B^R[1]$ and $\A^R[1]$, respectively.
In this way we find:
\bea
  \B^R_{m_1m_2}[1] & = & \frac{1}{(4\pi^2)^2} 
  \left[\frac{m_1 K_1(m_1 x) m_2 K_1(m_2 x)}{x^2}\right]^R \nn
  & = & 
  \frac{1}{32\pi^4} \left\{ \frac{m_1m_2}{m_1+m_2}
  [\Box-(m_1+m_2)^2]
  \frac{K_0(m_1 x) K_1(m_2 x) + K_0(m_2 x) K_1(m_1 x)}{x} 
  \right. \nn
  && \left. 
  \mbox{} + 
  2 \pi^2 \left( \log \frac{\bar{M}^2}{m_1m_2}
  + \frac{m_1-m_2}{m_1+m_2} \log \frac{m_2}{m_1}\right)
  \delta(x)  \right\}\, ,
\label{BRm1m2}
\eea
where $\bar{M}=2M/\gamma_E$ and $\gamma_E= 1.781 \dots$ is Euler's constant.
The massive one-point function $\A^R_m[1]$ is determined (see 
Appendix~\ref{Appren})
from $\A^R[1]$ and $\B^R_{m_1 m_2}$:
\be
  \A^R_m[1]=\frac{1}{16\pi^2} m^2 (1-\log \frac{\bar{M}^2}{m^2})
  \delta(x) \, .
\label{ARm}
\ee
The renormalization of the remaining basic functions is obtained
from $\B_{m_1m_2}^R[1]$ and
$\A_m[1]^R$ ($\B^R[1]$ and $\A^R[1]$ in the massless case)
by recurrence relations based on rules~\ref{R2}, \ref{R3}
and \ref{R4}. These recurrence relations follow from
the Leibnitz rule and two operations: {\em point separation} and 
{\em point contraction}.
Point separation~\cite{g2} allows to relate a generic renormalized
basic function with $n$ propagators 
and $r$ derivatives, $\F^{(n) R}_{m_1 \dots m_n}[\OO^{(r)}]$,
to renormalized basic functions 
with $n+1$ propagators and $r$, $r+1$ and $r+2$ 
derivatives. Using the rules,
\bea
  \lefteqn{\F^{(n) R}_{m_1 \dots m_n}[\OO](z_1,\dots,z_{n-1}) 
  \delta(z_n)} \nn
  & = & 
  \left[\prop_{m_1}(z_1) \dots \prop_{m_{n-1}}(z_{n-1})
  \OO^{z_1} \prop_{m_n}(z_1+ \cdots +z_{n-1}) \delta(z_n) \right]^R 
  \nn
  & = &  
  \left[\prop_{m_1}(z_1) \dots \prop_{m_{n-1}}(z_{n-1})
  \OO^{z_1} \prop_{m_n}(z_1+ \cdots +z_{n-1}+z_n) \delta(z_n) 
  \right]^R \nn
  & = & 
  - \left[\prop_{m_1}(z_1) \dots \prop_{m_{n-1}}(z_{n-1})
  \OO^{z_1} \prop_{m_n}(z_1+ \cdots +z_{n-1}+z_n) (\Box^{z_n} 
  - m_{n+1}^2) \prop_{m_{n+1}}(z_n) \right]^R \nn
  & = &
  (m_{n+1}^2 - \Box^{z_n}) 
  \F^{(n+1) R}_{m_1 \dots m_{n-1} m_{n+1} m_n}[\OO](z_1,\dots,z_n) 
  \nn
  && \mbox{} +
  2\d_\rho^{z_n} 
  \F^{(n+1) R}_{m_1 \dots m_{n-1} m_{n+1} m_n}
  [\OO \d_\rho](z_1,\dots,z_n) \nn
  && \mbox{} - 
  \F^{(n+1) R}_{m_1 \dots m_{n-1} m_{n+1} m_n}[\OO \Box](z_1,\dots,z_n)\, .
\label{pointseparation}
\eea
On the other hand, point contraction relates a renormalized basic
function with a d'alambertian and $r$ derivatives to
renormalized basic functions with one less propagator 
and $0,1,\dots,r$ derivatives:
\bea
  \lefteqn{\F^{(n+1) R}_{m_1 \dots m_{n+1}}[\OO \Box](z_1,\dots,z_n)} 
  \nn
  & = & 
  \left[\prop_{m_1}(z_1) \dots \prop_{m_n}(z_n) \OO^{z_1}
  \Box^{z_1} \prop_{m_{n+1}}(z_1 + \cdots + z_n) \right]^R \nn
  & = &
  - \left[ \prop_{m_1}(z_1) \dots \prop_{m_n}(z_n) \OO^{z_1}
  \left( \delta(z_1 + \cdots + z_n) - 
  m_{n+1}^2 \prop_{m_{n+1}}(z_1 + \cdots + z_n)\right) 
  \right]^R \nn
  & = &
  \sum_i \OO_i^{z_n} 
  \left( \F^{(n)R}_{m_1 \dots m_n}[\OO^\prime_i]
  (z_1,\dots,z_{n-1}) \delta(z_1+ \cdots + z_n) \right) \nn
  && \mbox{} +
  m_{n+1}^2 \F^{(n+1)R}_{m_1 \dots m_{n+1}}[\OO](z_1,\dots,z_n)\, .
\label{pointcontraction}
\eea
$\OO_i$ ($\OO^\prime_i$) is a differential operator with 
$r_i$ ($r-r_i$) derivatives, 
$0 \leq r_i \leq r$. 
Observe that point contraction and point separation are not
inverse operations of each other due to the different 
delta functions in the \lhs\ of Eq.~(\ref{pointseparation}) and
in the \rhs\ of Eq.~(\ref{pointcontraction}). This allows to 
obtain non-trivial information when Eq.~(\ref{pointcontraction})
is inserted in the last term of Eq.~(\ref{pointseparation}). 
In Appendix~\ref{Appren} we show how the combined use of the 
operations just described fixes the renormalization of all
the singular basic functions of Table~1.

\section{Renormalization of Scalar QED}

In this section we apply the procedure of CDR 
to Scalar QED and verify that the Ward identities among
renormalized Green functions are satisfied. 
The presence of derivative couplings makes
Scalar QED  the simplest renormalizable  theory that 
contains all the singular basic functions in Table~1. 
In this context, the techniques described above can be fully
illustrated. We calculate to one loop the
1PI Green functions of two, three and
four points, which are the only ones that require
renormalization. 
The one-point Green functions directly vanish.
In the following, after giving the
lagrangian and the Feynman rules, we briefly discuss the 
renormalization of the six 1PI Green functions that contain
singular graphs, namely the vacuum polarization,
the scalar selfenergy, the photon-scalar-scalar 
interaction vertex, 
and the three kinds of four particle vertices: photon-photon, 
photon-scalar and scalar-scalar scattering,
and the corresponding Ward identities. We also
check that the one-loop $\beta$-functions 
and anomalous dimensions of Scalar QED are recovered.

\begin{figure}
\epsfxsize=13cm
\begin{center}
\epsfbox{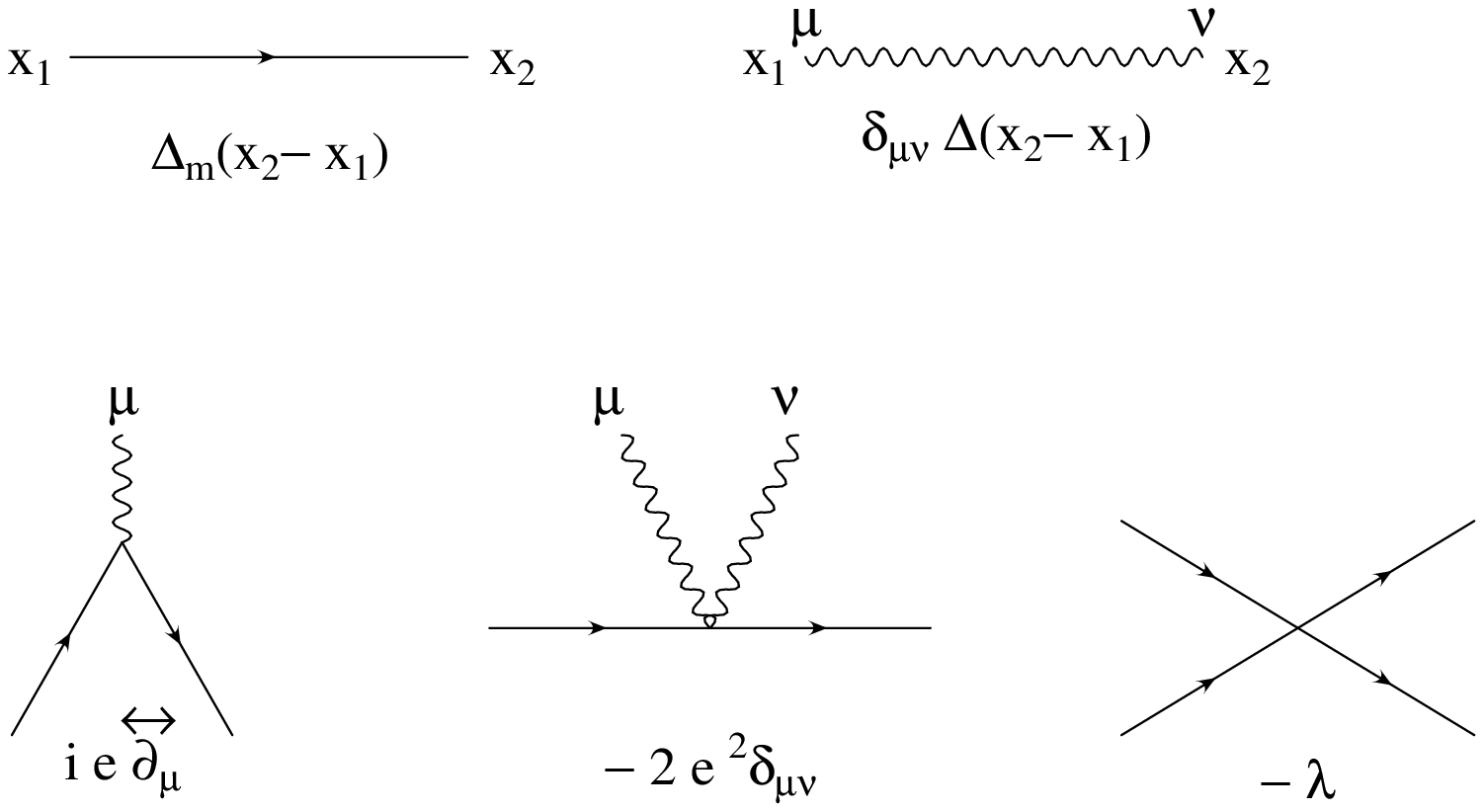}
\end{center}
\caption{Feynman rules of Scalar QED. Wavy lines correspond
to photons and solid lines to scalars. In the trilinear
coupling, $\stackrel{\leftrightarrow}{\d}_\mu =
\stackrel{\rightarrow}{\d}_\mu -
\stackrel{\leftarrow}{\d}_\mu$, with 
$\stackrel{\rightarrow}{\d}_\mu$ 
($\stackrel{\leftarrow}{\d}_\mu$) acting on the
incoming (outgoing) scalar.}
\end{figure}
The complete lagrangian for Scalar QED in euclidean space 
and in the Feynman gauge is
\be
{\mathcal L}_{\it E}  =  \frac{1}{4}F_{\mu\nu}F_{\mu\nu}+
                  \frac{1}{2}(\d_\mu A_\mu)^2+
                  (\d_\mu - ieA_\mu)\phi^{\dagger}(\d_\mu + 
                  ieA_\mu)\phi 
                  +m^2 \phi^\dagger \phi
                  + \frac{\lambda}{4}(\phi^\dagger \phi)^2   \, .
\label{lagra}
\ee
The corresponding  Feynman rules are depicted in Fig.1.

\begin{figure}
\begin{center}
\epsfxsize=13cm
\epsfbox{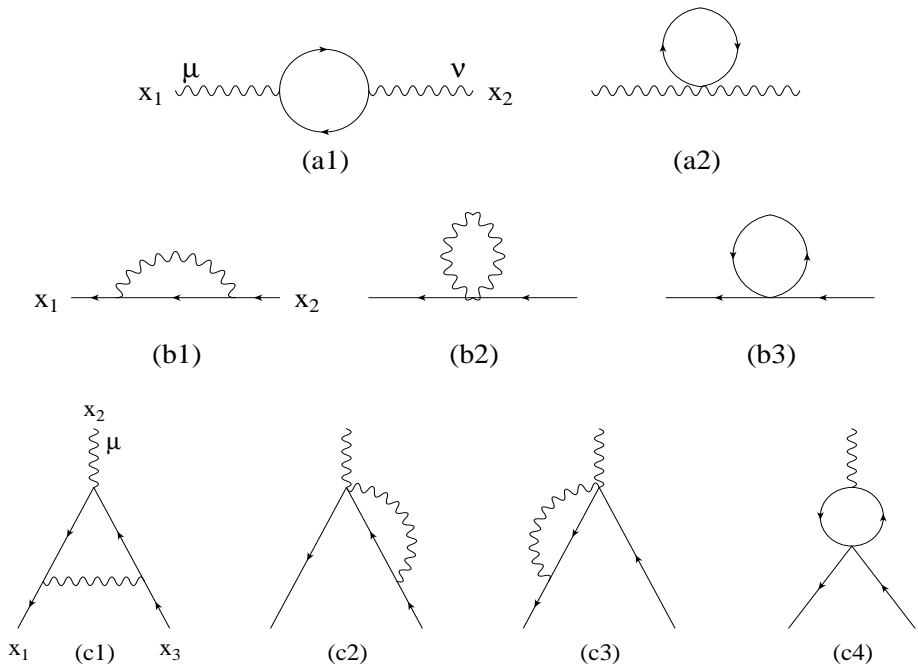}
\end{center}
\caption{Feynman diagrams contributing to
the vacuum polarization (a), scalar selfenergy (b) and
photon-scalar-scalar vertex (c).}
\end{figure}
\begin{figure}[h]
\begin{center}
\epsfxsize=14cm
\epsfbox{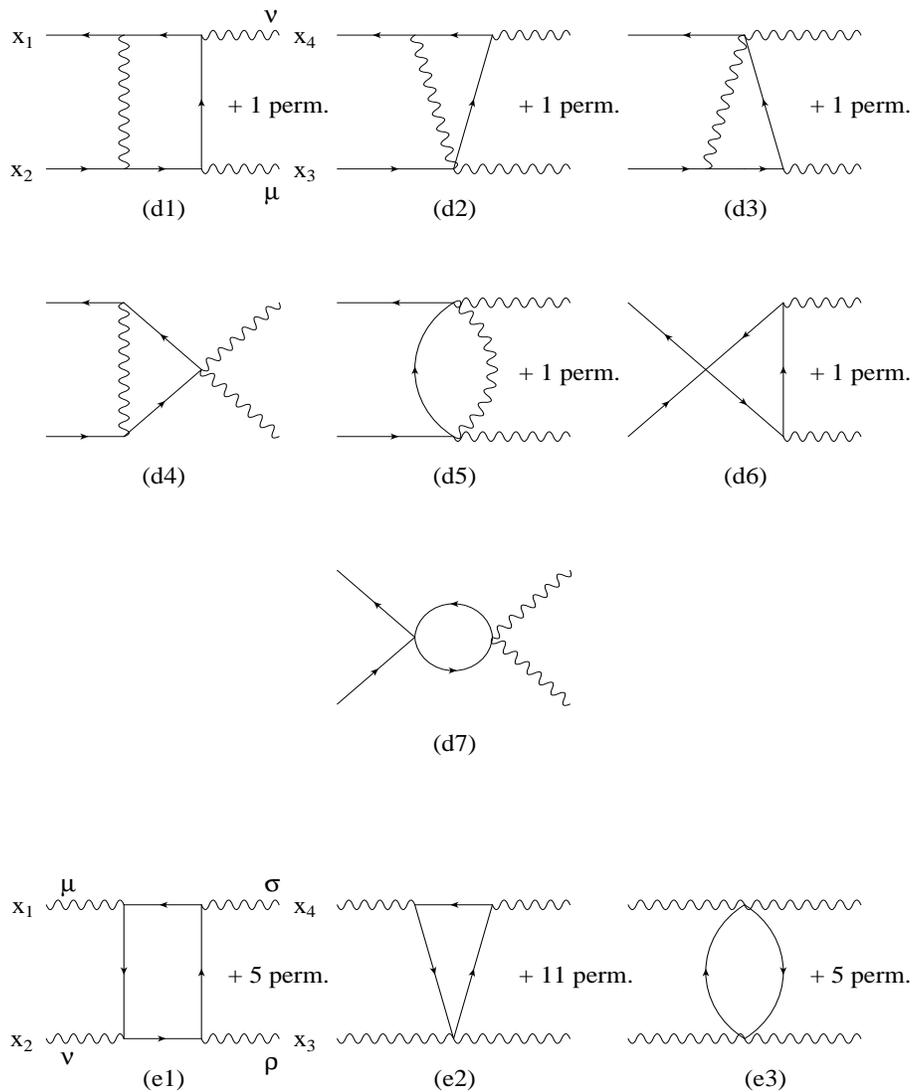}
\end{center}
\caption{Feynman diagrams contributing to 
the photon-scalar (d) and photon-photon (e) scattering. 
The permutations take into account diagrams with opposite
charge flow.}
\end{figure}
\begin{figure}[h]
\begin{center}
\epsfxsize=14cm
\epsfbox{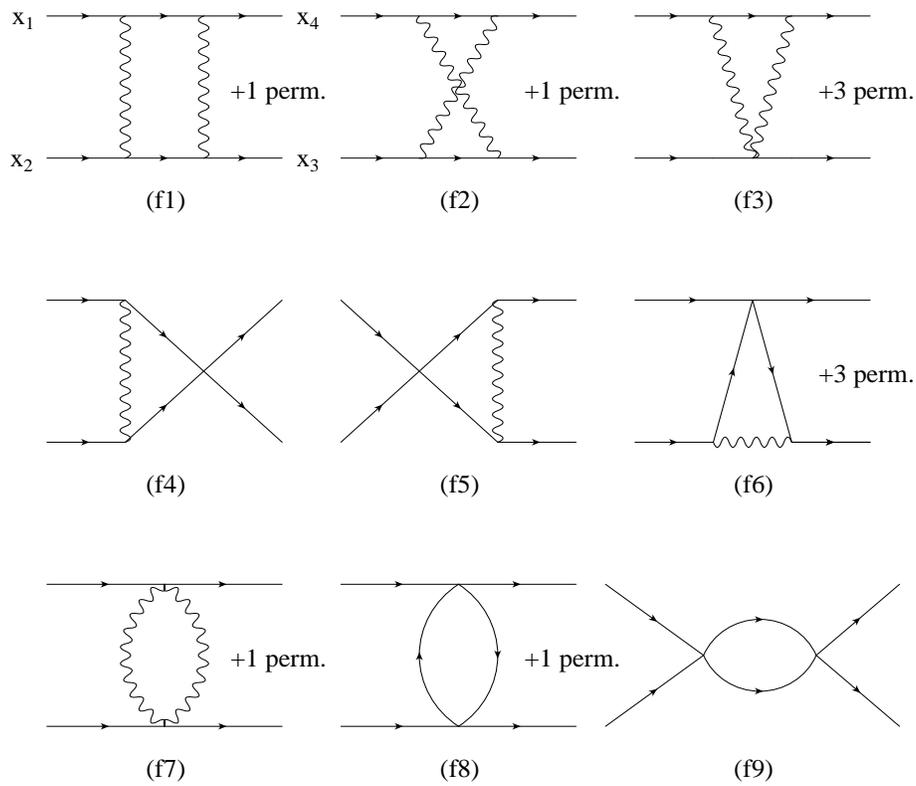}
\end{center}
\caption{Feynman diagrams contributing
to the scalar-scalar scattering (f).}
\end{figure}
\clearpage
\subsection{Vacuum polarization}
The two Feynman graphs contributing to the vacuum polarization, a1 and
a2, are depicted
in Fig.~2. Their contributions are
\bea
  \Pi_{(a)\mu\nu}(x)&=& -e^2\propm(x)
    \stackrel{\leftrightarrow}{\d}_\mu
    \stackrel{\leftrightarrow}{\d}_\nu \propm (x) \, , \nn
  \Pi_{(b)\mu\nu}(x)&=& -2e^2\delta_{\mu\nu}\propm (x)
    \delta (x) \, ,
\label{pol1}
\eea
where $x=x_1-x_2$ is the coordinate difference.
In the following, unless otherwise stated, we shall use 
the shifted variables defined 
in the text. In terms of basic functions the  two
contributions read
\bea
  \Pi_{(a)\mu\nu}(x)&=& -e^2(4\B_{mm}[\d_\mu\d_\nu]-
    \d_\mu\d_\nu\B_{mm}[1]) \, ,\nn 
  \Pi_{(b)\mu\nu}(x)&=& -2e^2\delta_{\mu\nu}\A_m \, .   
\label{pol2}
\eea
Using the expressions of the renormalized functions in Table~3, 
the sum is~\cite{CDR}
\bea
  \Pi_{\mu\nu}^R(x) & = & - \frac{e^2}{(4\pi^2)^2} 
    (\d_\mu\d_\nu - \delta_{\mu\nu} \Box) \left[ 
    \frac{1}{6} (\Box - 4m^2) (\frac{m K_0(mx) K_1(mx)}{x} 
    \right. \nn
    && \left. \mbox{} +  m^2 (K_0^2(mx)-K_1^2(mx))) + 
    \frac{1}{3} \pi^2 (\log \frac{\bar{M}^2}{m^2} - \frac{4}{3}) 
    \delta(x) \right]  \, ,
\label{polren}
\eea
which is transverse, as required by the Ward identity
\be
  \d_\mu^x \Pi_{\mu\nu}^R(x)  =  0 \, .
\label{trans2}
\ee
The scale-dependent part is
\be
  M\frac{\d}{\d \,M} \Pi_{\mu\nu}^R(x)  =  
    -\frac{e^2}{24\pi^2}
    \left(\d_{\mu}\d_{\nu}-\delta_{\mu\nu}\Box\right)\delta(x)\, .
  \label{derivative1}
\ee

\subsection{Scalar selfenergy}
The one--loop scalar selfenergy is obtained from graphs
b1, b2 and b3 in Fig.~2:
\bea
  \Sigma_{(a)}(x)&=&
    -e^2 \left[\Box \left(\prop(x) \propm(x)\right) + 
    2\d_\alpha \left(\prop(x) \d_\alpha \propm(x)\right) 
    \right. \nn
    && \mbox{} \left. + \prop(x)\Box\propm(x) \right] \, , \\
  \Sigma_{(b)}(x)&=&-4e^2\prop (x)\delta (x) \, , \\ 
  \Sigma_{(c)}(x)&=&-\lambda\propm (x)\delta (x) \, ,
\eea
where derivatives acting on the external scalars have been
changed by minus derivatives acting on the whole amputated
expression. In terms of basic functions, 
the total renormalized selfenergy is
\be
  \Sigma^R(x)  =  -e^2 \left(\Box \B_{0m}^R[1]+
  2\d_\rho \B_{0m}^R[\d_\rho] + \B_{0m}^R[\Box]
  +4 \A^R[1] \right) - \lambda \A_m^R[1] \, .
\ee
The explicit expression can be obtained using Tables~2 and~3. 
The scale-dependent part reads
\be
  M\frac{\d}{\d \,M} \Sigma^R(x)  =   \left( 
  -\frac{e^2}{8\pi^2}
  ( 2\Box + m^2) +
  \frac{\lambda}{8\pi^2}m^2 \right) \delta(x) \, .
\label{derivative2} 
\ee
\subsection{Photon-scalar-scalar vertex}
The contributing diagrams, c1-c4, are shown in 
Fig.~2. Their renormalized
expressions in terms of 
basic functions are
\bea
V_{(a)\mu}^R & = &
  - ie^3 \left\{-2\T_{mm0}^R[\Box\d_\mu]+(\d^x_\mu+\d^y_\mu) 
  \T_{mm0}^R[\Box]+4(\d^x_\alpha+\d^y_\alpha)  
  \T_{mm0}^R[\d_\mu\d_\alpha] \right. \nn 
  & & \mbox{} 
  -8\d^x \cdot\d^y \T_{mm0}[\d_\mu]-2(\d^x_\mu+\d^y_\mu)  
  (\d^x_\alpha+\d^y_\alpha)\T_{mm0}[\d_\alpha] \nn
  & & \left. \mbox{} +
  4(\d^x_\mu+\d^y_\mu)\d^x \cdot \d^y \T[1]\right\} \, , 
  \label{Vamu} \\
V_{(b)\mu}^R & = &
  - 2ie^3 \left\{ \d_\mu^y \B_{0m}^R[1](y) 
  + \B_{0m}^R[\d_\mu](y) \right\} \delta(x) \, ,\\  
V_{(c)\mu}^R & = &
  - 2ie^3 \left\{ \d_\mu^x \B_{0m}^R[1](x) 
  + \B_{0m}^R[\d_\mu](x) \right\} \delta(y) \, ,
  \label{Vcmu} \\  
V_{(d)\mu}^R & = &0. 
\eea
Graph c4, which is proportional to $\lambda$, 
vanishes directly because of charge conjugation.
The renormalized vertex results from the sum
of Eqs.~(\ref{Vamu}-\ref{Vcmu}) and the use
of Tables~3 and~4.
The scale-dependent part is
\be
  M\frac{\d}{\d \,M} V_{\mu}^R(x,y) = -i\frac{e^3}{4\pi^2}
    ( \d_{\mu}^x + \d_{\mu}^y)\delta(x)\delta(y) \, .
  \label{derivative3}
\ee
The renormalized vertex function and selfenergy fulfil
the Ward identity
\be
 (\d_\mu^x-\d_\mu^y) V_\mu^R(x,y) = ie(\Sigma^R(x)\delta(y)-
   \Sigma(y)^R\delta(x))\, , 
 \label{ward12}
\ee
analogous to Eq.~(\ref{WI}) in QED.
This can be checked integrating with any test function
$\phi (x,y)$. In
particular, $\phi (x,y)=e^{i p \cdot x}$ leads to simple
expressions for both sides of the identity~\cite{polonia}.
Observe that the part of the Ward identity proportional to
$\lambda$ (coming from diagrams~b3 and~c4) is trivially satisfied, 
for both sides vanish.
In general, the diagrams proportional to $\lambda$ form a gauge
invariant subset.

\subsection{Four-point functions}
The diagrams contributing to the photon-scalar ($V_{\mu\nu}$),
d1-d7, and
photon-photon ($V_{\mu\nu\rho\sigma}$), e1-e3, and 
to the scalar-scalar
($V$), f1-f9, scattering are depicted in
Figs.~3 and~4, respectively.
As in previous cases, the renormalized Green functions
are obtained expanding in basic functions and
substituting these by their renormalized expressions
in Tables~2-4. The final results are too lengthy to be reported
here. We have calculated them with a symbolic program
based on {\em Mathematica}.
The scale-dependent parts are readily obtained. For the 
photon-photon-scalar-scalar vertex we have
\be
  M\frac{\d}{\d \,M} V_{\mu\nu}^R(x,y,z) =  \delta_{\mu\nu}
    \frac{e^4}{2\pi^2} \delta(x)\delta(y)\delta(z) \, .
  \label{derivative4}
\ee
Notice that there is no $\lambda$-dependent part.
Actually, Eqs.~(\ref{derivative3}) and
(\ref{derivative4}) follow from gauge invariance and
Eq.~(\ref{derivative1}).
The scale-dependent part of the four-photon vertex
cancels when all diagams are summed, as required by
renormalizability.
Finally, the scale-dependent part of the
four-scalar vertex is
\be
  M\frac{\d}{\d \,M} V(x,y,z) = 
    \frac{1}{16\pi^2}(24 e^4-4e^2\lambda+5\lambda^2) 
    \delta(x)\delta(y)\delta(z).  
  \label{derivative5}
\ee
Observe that there is a $\lambda$-independent piece, reflecting
the well-known fact that a quartic scalar self-coupling 
is needed in the lagrangian for multiplicative renormalizability.
The Ward identities for the
four-point functions are more transparent in terms of the
original external points:
\bea
  &&\d_\mu^{x_3}  V_{\mu\nu}^R (x_1,x_2,x_3,x_4)  =    
    ie  V_\nu^R (x_1,x_4,x_2)
    (\delta(x_1-x_3)-\delta(x_2-x_3))\, , 
    \label{ward22} \\    
  &&\d_{\mu}^{x_1}  V_{\mu\nu\rho\sigma}^R (x_1,x_2,x_3,x_4) 
    = 0 \, , \label{ward32} 
\eea
and the identities related to Eqs.~(\ref{ward22}) 
and~(\ref{ward32}) by Bose symmetry.
There is no Ward identity relating $V$ to other
singular Green functions.
In terms of the shifted variables,
Eqs.~(\ref{ward22}) and~(\ref{ward32}) read
\bea
  && (\d_\mu^z-\d_\mu^y) V_{\mu\nu}^R (x,y,z) =
    ie V_\nu^R (x+y+z,-y-z) [\delta(x+y)-\delta(y)] \, ,
    \label{ward22xyz} \\
  && \d_{\mu}^{x}  V_{\mu\nu\rho\sigma}^R (x,y,z) 
    = 0 \, . \label{ward32xyz} 
\eea
The scale-dependent parts of $V_{\mu\nu}^R$,
$V_\mu^R$ and $V_{\mu\nu\rho\sigma}^R$ trivially
satisfy Eqs.~(\ref{ward22xyz}) and~(\ref{ward32xyz}).
The merit of CDR is to give also the correct
scale-independent parts of the Ward identities. 
We have checked numerically that this is indeed the case,
supplementing the program used for the calculation
of the renormalized Green functions with the
{\em FF} package~\cite{FF} ({\em FF} works in 
dimensional regularization, but we have only used
it for finite functions).

\subsection{Renormalization group} 
To conclude this section we compute the first order anomalous 
dimensions and $\beta$-functions. 
The renormalization group equation for any renormalized 
amplitude reads
\be
\left( M\frac{\d}{\d \,M} + \beta_e \frac{\d}{\d \,e}+
\beta_\lambda \frac{\d}{\d \,\lambda}+  \gamma_m m 
\frac{\d}{\d \, m} -n_{\phi}\gamma_{\phi}
-n_{A}\gamma_A \right) \Gamma^{R(n_\phi,n_A)} =0 \, ,
\label{rge}
\ee
where $\gamma_{\phi,A}$ are the anomalous dimensions 
of the corresponding fields
and  $\Gamma^{R(n_\phi,n_A)}$ is the 1PI renormalized 
amplitude of $n_\phi$ scalar and $n_A$ photon
fields\footnote{
The only effect of the gauge parameter dependence is 
to consistently cancel the tree-level 
longitudinal term in the equation for the vacuum polarization.}.
The tree level part of the $\Gamma$ functions can be read off
from the Feynman rules in Fig.~1 and the 
scale-dependence of the one--loop amplitudes is
given in Eqs.~(\ref{derivative1}), (\ref{derivative2}),
(\ref{derivative3}), (\ref{derivative4}) and 
(\ref{derivative5}).
Using the renormalization group equations (\ref{rge}) 
for the relevant amplitudes
one obtains
\bea
\gamma_A        &=&   \frac{e^2}{48\pi^2} \, , \\
\gamma_\phi     &=&    -\frac{e^2}{8\pi^2} \, , \\
\gamma_m        &=&    \frac{\lambda-3e^2}{16\pi^2} \, , \\
\beta_e         &=&    \frac{e^3}{48\pi^2} \, , \\
\beta_\lambda   &=&    \frac{24e^4-12e^2\lambda+
                       5\lambda^2}{16\pi^2} \, .
\eea
Notice that $\beta_e=e\gamma_A$, as dictated by gauge invariance.

\section{Conclusions}
Constrained differential renormalization is
a scheme of differential renormalization which
does not introduce the ambiguities
inherent to formal manipulations of singular expressions. 
At the moment it has
only been developed at the one-loop level, although
in principle it could be extended to
higher orders. Such extension looks rather
involved and is under study.
In this paper we provide the techniques and explicit 
expressions that allow to perform any one-loop calculation
in a four dimensional renormalizable theory (in the Feynman
gauge) with this method. The whole procedure has been
implemented in a computer package that performs all
operations automatically in momentum space. This package is
available and its description can be found in Ref.~\cite{program}.
Constrained differential renormalization
respects the (one-loop) Ward identities of 
gauge invariance.
Here, we have shown it for Scalar QED, which contains
all the singular functions appearing
in renormalizable theories. In Ref.~\cite{CDR} it was verified that
the Ward identities in QED are satisfied and 
the correct triangular anomaly recovered (a more complete 
discussion of anomalies will be presented elsewhere). That
non-abelian gauge invariance is preserved is 
explicitly shown in Ref.~\cite{QCD}. 
The method also preserves supersymmetry in the
calculation of the supergravity corrections to the $g-2$ of a 
charged lepton~\cite{g2,SUSY}.

\section*{Acknowledgements}
We thank T. Hahn for providing
a helpful package linking {\em Mathematica} and {\em FF}. 
The pictures have been made with {\em FeynDiag}.
MPV thanks D.Z. Freedman for discussions and MIT for 
its hospitality.
This work has been supported by CICYT, under contract number
AEN96-1672, and by Junta de Andaluc\'{\i}a, FQM101.
RMT and MPV thank Ministerio de Educaci\'on y Cultura for
financial support.

\clearpage

\appendix

\section{Appendix}
\label{Appren}
In this appendix we derive the renormalized expressions of all the 
singular basic functions appearing in renormalizable theories.
Essentially, we use the techniques of point separation and
point contraction, although in the simplest cases some intermediate
steps can be avoided. For simplicity,
we deal first with massless functions and then discuss the minor
modifications of the process needed for the massive case.

\subsection{Massless basic functions}
Let us start with the one-point basic functions. We saw in the
text that rule~\ref{R1a} implied 
$\A^R[1]=a \Box \delta(x)$, with $a$ arbitrary in principle.
Consider now the expression
$\A^R[1] \delta(y)$. Using rule~\ref{R3},
\bea
  \A^R[1](x) \delta(y) & = & \left[\prop(x) \delta(x)\right]^R 
  \delta(y)
  = \left[ \prop(x) \delta(x) \delta(x+y) \right]^R  \nn
  & = & \A^R[1](x) \delta(x+y)\, . 
\eea
Integration on $x$ leads to the equation
\be
  \delta(y) \int {\mathrm d}^4 x \, \A^R[1](x) = \A^R[1](y) \, ,
\ee
which, using rule~\ref{R2}, requires $a=0$. Therefore, 
\be
   \A^R[1]= 0\, .
\ee
Similarly, locality and power counting imply that
$A^R[\d_\mu]=[\d_\mu \prop(x) \delta(x)]^R$ is of the form
\be
  \A^R[\d_\mu]=(a^\prime \Box + \mu^{\prime 2}) \d_\mu \delta(x)\, ,
\ee
and the same argument used before implies 
$a^\prime=\mu^\prime=0$, and hence
\be
  \A^R[\d_\mu]=0\, .
\ee
So, the massless one-point basic functions are zero in
CDR. (The same results are obtained with the general form of
point separation and point contraction.)

Consider now the two-point basic functions.
$\B^R[1]$ was already given in the text. 
$\B^R[\d_\mu]$ is easily obtained using the Leibnitz rule 
and the fact that the two propagators involved are identical:
\be
  \B^R[\d_\mu]  =  \left[ \prop(x) \d_\mu \prop(x) \right]^R \nn
  = 
  \d_\mu \left[\prop(x) \prop(x)\right]^R - 
  \left[ \d_\mu \prop(x) \prop(x) \right]^R 
\ee
and then
\be
  \B^R[\d_\mu]  =  \frac{1}{2} \d_\mu 
    \left[(\prop(x))^2\right]^R \nn
   =  \frac{1}{2} \d_\mu \B^R[1] \, .
\label{symmetricLeibnitz}
\ee
$\B^R[\Box]$ is directly obtained from $\A^R[1]$ using
rule~\ref{R4}:
\be
  \B^R[\Box]=\left[\prop(x)\Box\prop(x)\right]^R =
  -\left[\prop(x)\delta(x)\right]^R = -\A^R[1]=0 \, .
\label{BRBox}
\ee
This is the simplest example of point contraction.
The renormalization of $\B[\d_\mu\d_\nu]$ is more involved.
First, we notice that its most general
renormalized expression is
\be
  \B^R[\d_\mu\d_\nu] = 
  \frac{1}{3} (\d_\mu\d_\nu-\frac{1}{4}\delta_{\mu\nu}\Box)
  \B^R[1] +
  \frac{1}{16\pi^2} 
  [f \d_\mu\d_\nu + \delta_{\mu\nu}( g \Box + 
  \mu^{\prime\prime \, 2} )]\delta(x)\, ,
\label{genBRdd}
\ee
where the first term results from solving a differential
equation at $x \not = 0$ and
$f$, $g$ and $\mu^{\prime\prime}$ are arbitrary coefficients
with dimension 0, 0 and 2, respectively,
parametrizing the ambiguity in the local terms\footnote{
The dimensionful coefficient $\mu^{\prime\prime}$ is forbidden
from rule~\ref{R1a}, but we shall instead use 
recurrence
relations and the initial condition $\A^R[1]=0$ to prove 
that it vanishes, thus
showing that rule~\ref{R1a} is consistent with the
other rules.}.
Second, applying rule~\ref{R4} to the delta function in 
$\A[\d_\mu]$ we obtain
\be
  0 = \A^R[\d_\mu]=-\d_\mu B^R[\Box]+\B^R[\d_\mu \Box] 
  = \B^R[\d_\mu\Box]   \label{pepe} \, ,
\ee
and using the Leibnitz rule to move
the derivatives to the first (identical) propagator and
Eqs.~(\ref{symmetricLeibnitz}) and~(\ref{BRBox}),
\be
  \B^R[\d_\mu\Box] =
  - \frac{1}{2} \d_\mu \Box \B^R[1] + 2\d_\rho \B^R[\d_\mu\d_\rho]
  - \B^R[\d_\mu\Box] \label{juan} \, .
\ee
Inserting Eq.~(\ref{pepe}) 
into Eq.~(\ref{juan}), we find the consistency equation
\be
  -\frac{1}{2} \Box\d_\mu \B^R[1] + 
  2\d_\rho \B^R[\d_\mu\d_\rho] = 0 \, .
\label{BRdd}
\ee 
If we substitute Eq.~(\ref{genBRdd}) into Eq.~(\ref{BRdd})
we find $\mu^{\prime\prime}=0$ and $g=-f$. For the moment,
$f$ remains arbitrary. It will be fixed from the relation
between $\B^R[\d_\mu\d_\nu]$ and $\T^R[\d_\mu\d_\nu\d_\rho]$ later 
on.

The renormalization of three-point basic functions containing one
d'alambertian is directly obtained via point contraction, which
relates them to two-point functions already renormalized:
\bea
  \T^R[\Box] & = & -\B^R[1] \delta(x+y)\, , \\
  \T^R[\Box\d_\mu] & = & - \d_\mu^y (\B^R[1](x) \delta(x+y))
  - \B^R[\d_\mu](x) \delta(x+y) = \nn
  && \mbox{} - \frac{1}{2}
  (\d_\mu^x+\d_\mu^y) (\B^R[1] \delta(x+y)) \, .
  \label{TRdbox}
\eea
For T functions with more complex index structure we
apply point separation to the corresponding B function
with one less derivative. $\T^R[\d_\mu\d_\nu]$ can be decomposed
into a traceless and a trace part, plus a possible
ambiguous local term (symmetric under $\mu \leftrightarrow \nu$ and 
$x \leftrightarrow y$ and with the correct
dimension)~\cite{CDR}:
\be
  \T^R[\d_\mu\d_\nu] = \T[\d_\mu\d_\nu-\frac{1}{4}\delta_{\mu\nu}
  \Box] + \frac{1}{4} \delta_{\mu\nu} \T^R[\Box] +
  \frac{1}{64\pi^2} \, b \, \delta_{\mu\nu} \delta(x)\delta(y) \, .
\label{traceTdd}
\ee
The traceless part is finite and the trace has already been
renormalized, so we just have to fix the dimensionless
coefficient $b$. Separating points,
\be
  \B^R[\d_\mu](x)\delta(y) =
  -\Box^y \T[\d_\mu] + 2 \d_\rho^y \T^R[\d_\mu\d_\rho]
  - \T^R[\Box\d_\mu]\, .
\ee
The simplest way to solve this equation for $b$ 
is to integrate on $x$ (see Ref.~\cite{CDR}).
Integration on the ``separated'' point, $y$, leads to a 
tautology (this is a general fact). We obtain the value
$b=-\frac{1}{2}$.
The same process can be applied to $\T^R[\d_\mu\d_\nu\d_\rho]$.
Its trace-traceless decomposition is
\bea
  \T^R[\d_\mu\d_\nu\d_\rho] & = &
  \T[\d_\mu\d_\nu\d_\rho-\frac{1}{6}(\delta_{\mu\nu}
  \d_\rho+\delta_{\mu\rho}\d_\nu+\delta_{\nu\rho}\d_\mu)\Box] \nn
  & & \mbox{} + \frac{1}{12}
  \left( \delta_{\mu\nu}(\d_\rho^x+\d_\rho^y) + 
  \delta_{\mu\rho}(\d_\nu^x+\d_\nu^y) +
  \delta_{\nu\rho}(\d_\mu^x+\d_\mu^y) \right) \nn
  & & \mbox{} \times \left(-\B^R[1] \delta(x+y) +
  \frac{1}{16\pi^2} \, c \, \delta(x)\delta(y) \right) \, ,
\label{traceTddd}
\eea
where we have used Eq.~(\ref{TRdbox}).
Point separation of $\B^R[\d_\mu\d_\nu]$, followed by point 
contraction of the T functions with one d'alambertian, leads to
\bea
  \B^R[\d_\mu\d_\nu](x)\delta(y) & = &
  -\Box^y \T^R[\d_\mu\d_\nu] + 
  2 \d_\rho^y \T^R[\d_\mu\d_\nu\d_\rho] \nn
  && \mbox{} + (\d_\mu^x\d_\nu^y+\d_\mu^y\d_\nu^x)
  \left(\B^R[1](x)\delta(x+y)\right) \nn
  && \mbox{} + 
  \B^R[\d_\mu \d_\nu](x)\delta(x+y)\, .
\eea
Integrating on $x$ after inserting Eqs.~(\ref{traceTdd}) 
and~(\ref{traceTddd}), we get 
\bea
  0 & = & -\Box^y \int {\mathrm d}^4 x \, 
  \T[\d_\mu\d_\nu - \frac{1}{4}\delta_{\mu\nu}\Box] +
  2\d_\rho^y \int {\mathrm d}^4 x \,
  \T[\d_\mu\d_\nu\d_\rho-\frac{1}{6}(\delta_{\mu\nu}
  \d_\rho+\delta_{\mu\rho}\d_\nu+\delta_{\nu\rho}\d_\mu)\Box] \nn
  && \mbox{} + \frac{1}{16\pi^2} \left[
  (\frac{1}{8} + \frac{c}{6} - f)\delta_{\mu\nu} \Box^y
  + (\frac{c}{3} + f) \d_\mu^y\d_\nu^y) \right] \delta(y)\, ,
\eea
which provides two equations (one for each tensor structure) that
determine the coefficients $f$ and $c$.
Calculating the (finite) integrals we find
$f=\frac{1}{18}$ and $c=-\frac{1}{3}$.

For four-point basic functions, which are at most logarithmically
singular, the situation is similar. Q functions with
at least one d'alambertian are directly contracted into
T functions:
\bea
  \Q^R[\Box\Box] & = & - \Box^z \left(\T[1]\delta(x+y+z)\right)
  -2\d_\rho^z \left(\T[\d_\rho](x,y)\delta(x+y+z)\right) \nn
  && \mbox{} - \T^R[\Box](x,y)  \delta(x+y+z)\, , \\
  \Q^R[\Box\d_\mu\d_\nu] & = & 
  -\d_\mu^z\d_\nu^z \left(\T[1]\delta(x+y+z)\right)
  - \d_\mu^z \left(\T[\d_\nu](x,y)\delta(x+y+z)\right) \nn
  && \mbox{} -\d_\nu^z \left(\T[\d_\mu](x,y)\delta(x+y+z)\right) \nn
  && \mbox{} - \T^R[\d_\mu\d_\nu](x,y) \delta(x+y+z)\, .
\eea
Finally, the renormalization of $\Q[\d_\mu\d_\nu\d_\rho\d_\sigma]$
requires again the combined use of point separation and point
contraction. The trace-traceless decomposition is not unique
in this case. The simplest possibility is
\bea
  \Q^R[\d_\mu\d_\nu\d_\rho\d_\sigma] & = &
  \Q[\d_\mu\d_\nu\d_\rho\d_\sigma-\frac{1}{24}
  (\delta_{\mu\nu}\delta_{\rho\sigma} +
  \delta_{\mu\rho}\delta_{\nu\sigma} +
  \delta_{\mu\sigma}\delta_{\nu\rho})\Box\Box] \nn
  &&\mbox{} + \frac{1}{24} (\delta_{\mu\nu}\delta_{\rho\sigma} +
  \delta_{\mu\rho}\delta_{\nu\sigma} +
  \delta_{\mu\sigma}\delta_{\nu\rho}) \nn
  && \mbox{} \times (\Q^R[\Box\Box] 
  + \frac{1}{16\pi^2} \,d \, \delta(x)\delta(y)\delta(z))\, .
\eea
Point separation of $\T^R[\d_\mu\d_\nu\d_\rho]$ gives
\be
  \T^R[\d_\mu\d_\nu\d_\rho](x,y)\delta(z) =
  -\Box^z \Q[\d_\mu\d_\nu\d_\rho] + 
  2\d_\sigma^z \Q^R[\d_\mu\d_\nu\d_\rho\d_\sigma]
  - \Q^R[\Box\d_\mu\d_\nu\d_\rho]\, ,
\label{traceQRdddd}
\ee
and point contraction of $\Q^R[\Box\d_\mu\d_\nu\d_\rho]$
plus integration on $x,y$ give 
the equation
\bea
  0 & = & -\Box^z \int {\mathrm d}^4 x {\mathrm d}^4 y \,
  \Q[\d_\mu\d_\nu\d_\rho] + 
  2 \d_\sigma^z \int {\mathrm d}^4 x {\mathrm d}^4 y \,
  \Q^R[\d_\mu\d_\nu\d_\rho\d_\sigma] \nn
  && \mbox{} + \int {\mathrm d}^4 y \, 
  \T^R[\d_\mu\d_\nu\d_\rho](y,z)\, .
\eea
Inserting Eq.~(\ref{traceQRdddd}) and the expression obtained
for $\T^R[\d_\mu\d_\nu\d_\rho]$, and calculating the integrals,
we find $d=\frac{5}{6}$, what completes the
renormalization of $\Q[\d_\mu\d_\nu\d_\rho\d_\sigma]$.

\subsection{Massive basic functions}
The massive case is analogous to the massless one. The degree
of the leading singularities is not changed by the
inclusion of masses.
Point separation and
point contraction are used in the same manner, although now 
the resulting
expressions contain additional terms proportional to the masses
(see Eqs~(\ref{pointseparation}) and~(\ref{pointcontraction})).
In order to use (when possible) symmetrically the Leibnitz rule, 
we decompose the basic functions into two pieces, 
one symmetric in the masses---that 
is treated as in the massless case---and one 
antisymmetric---that is less singular.
 
Let us illustrate the new features of the 
general massive case. 
The renormalization of 
$\A_m[1]=\prop_m(x)\delta(x)$,
is a bit more involved than that of $\A[1]$.
The renormalized expression must be of the form
\be 
  \A^R_m[1]= (\tilde{a} \Box + \tilde{\mu}^2) \delta(x) \, .
\ee
The same argument used for $\A^R[1]$ forbids the 
term proportional to $\Box \delta(x)$. On the other hand,
using rule~\ref{R4} and moving derivatives to
the massive propagator we get
\bea
  \A^R_m[1] &=& - \left[\prop_m(x) \Box \prop(x)\right]^R \nn
  &=& -\Box \B^R_{0m}[1] + 
  2 \d_\rho \B^R_{0m}[\d_\rho] + \A^R[1] - m^2 \B^R_{0m}[1]\, .
\eea
Then, integrating this equality (with rule~\ref{R2}),
\be
  \tilde{\mu}^2 = -m^2 \int {\mathrm d}^4 x \, \B^R_{0m}[1].
\ee
So, we only have to integrate Eq.~(\ref{BRm1m2}) and take the limit
$m_1 \rightarrow 0$. (The integrals of expressions containing
modified Bessel functions are easily performed with the
techniques described in Appendix~C of Ref.~\cite{g2}.)
The result is given in Eq.~(\ref{ARm}). Observe that 
$\tilde{\mu}$ is proportional to $m$, as it should for consistency
with rule~\ref{R1a}.

$\A_m^R[\d_\mu]$ is renormalized analogously to $\A^R[\d_\mu]$.
$\B_{m_1m_2}^R[1]$ is given in the text, whereas
$\B_{m_1m_2}^R[\d_\mu]$ is obtained generalizing 
Eq.~(\ref{symmetricLeibnitz}) with the
decomposition mentioned above:
\bea
  \lefteqn{\B_{m_1m_2}^R[\d_\mu] =} \nn
    & & \frac{1}{2} \left(\B_{m_1m_2}^R[\d_\mu] +
    \B_{m_2m_1}^R[\d_\mu] \right) + 
    \frac{1}{2} \left(\B_{m_1m_2}[\d_\mu] -
    \B_{m_2m_1}[\d_\mu]\right) \nn
     & =  & \frac{1}{2} \d_\mu \B_{m_1m_2}^R[1] \nn
    & & \mbox{} + 
    \frac{1}{64\pi^4} \left[
    m_2^2 K_0(m_2x)\d_\mu \frac{K_1(m_1x)}{x}
    -m_1^2 K_0(m_1x) \d_\mu \frac{K_1(m_2x)}{x} \right] \, .
\eea
On the other hand, $\B_{m_1m_2}[\Box]$ as well as
T and Q functions are renormalized following the massless case
but including the terms proportional to the masses, \eg:
\be
  \B_{m_1m_2}^R[\Box] = m_2^2 \B_{m_1m_2}^R[1] - 
  \A_{m_1}^R[1] \, ,
\ee
to be compared to Eq.~(\ref{BRBox}).
Finally, $\B_{m_1m_2}[\d_\mu\d_\nu]$ requires some
work to solve the differential equation.  Afterwards, the local
terms are fixed analogously to the $\B[\d_\mu\d_\nu]$ case.
The final result is given in Table~3.

\section{Appendix}
\label{AppFourier}

In this appendix we give all the ingredients needed to obtain
the momentum space expressions of Green functions renormalized
with CDR. Since the renormalized functions are well-defined
distributions, they admit a finite Fourier transform without
any regulator.
The Fourier transform of a distribution $f(x_1,\dots,x_n)$,
where $x_i$ are four-dimensional variables, is
\be
  \hat{f}(p_1,\dots,p_n) =
  \int {\mathrm d}^4 x_1 \dots {\mathrm d}^4 x_n \,
  e^{i x_1 \cdot p_1} \dots e^{i x_n \cdot p_n}
  f(x_1,x_2,\dots,x_n) \, .
\ee
Due to rule~\ref{R2}, total derivatives directly yield
\be
  \d_\mu^{x_j} \rightarrow -i p_{j \, \mu} \, .
\ee
Delta functions give rise to reduced Fourier
transforms, with linear 
combinations of the original momenta as arguments.
For instance,
\be
  f(x_1,\dots,x_{n-1}) \delta(x_{n-1}+x_n) \rightarrow
  \hat{f}(p_1,\dots,p_{n-1}-p_n) \, .
\ee
Therefore, one only needs the Fourier transforms 
of the renormalized basic functions. 
We shall perform Fourier transforms in the shifted variables $z_i$.
The Fourier transform of basic functions that are directly
finite without renormalization is simply a convolution in
momentum space~\cite{polonia}:
\bea
  \lefteqn{\hat{\F}^{(n)}_{m_1\dots m_n}[\OO](p_1,\dots,p_{n-1})} 
  \nn
  & = & \int \frac{{\mathrm d}^4 k}{(2\pi)^4} \,
  \frac{\hat{\OO}(k)}{[k^2+m_n^2] [(k-p_1)^2+m_1^2] \dots
  [(k-p_{n-1})^2+m_{n-1}^2]} \nn
  & \equiv & \I^{(n)}_{m_nm_1\dots m_{n-1}}[\hat{\OO}]
  (p_1,\dots,p_{n-1}) \, .
\label{finiteintegral}
\eea
The integrals $\I^{(n)}$ appear in usual one-loop calculations
in momentum space and can be calculated with
standard techniques. 

To Fourier transform the renormalized basic
functions of Tables~2, 3 and~4 we only need
(besides Eq.~(\ref{finiteintegral})) the following
integrals:
\bea
  \int {\mathrm d}^4 x \, e^{i x\cdot p} \frac{\log x^2M^2}{x^2} 
  & = & \frac{4\pi^2}{p^2} \log{\bar{M}^2}{p^2} \, , \\
  \int {\mathrm d}^4 x \, e^{i x\cdot p} 
  \frac{m K_1(mx)}{x} 
  & = &  4\pi^2 \frac{1}{p^2+m^2} \, , \\
  \int {\mathrm d}^4 x \, e^{i x\cdot p} K_0(mx)
  & = & 8\pi^2 \frac{1}{(p^2+m^2)^2} \, , \\
  \int {\mathrm d}^4 x \, e^{i x\cdot p} 
  K_0(m_1 x) \OO \frac{m_2 K_1(m_2 x)}{x} 
  & = & 32 \pi^4 \, \I^{(3)}_{m_2m_1m_1}[\hat{\OO}](p,p) \, , \\
  \int {\mathrm d}^4 x \, e^{i x\cdot p} 
  \frac{m_1 K_1(m_1 x)}{x} \OO K_0(m_2 x)
  & = & 32\pi^4 \, \I^{(3)}_{m_2m_2m_1}[\hat{\OO}](0,p) \, .
\eea
The Fourier transforms involving modified Bessel
functions are easily obtained using recurrence relations
among them~\cite{g2,Abramowitz}. The
recurrence relations include delta 
functions to make them valid at the origin.

In Tables~5, 6 and~7 we collect
the Fourier transforms of the basic functions in
Tables~2, 3 and~4, respectively.
These basic functions in (euclidean) momentum space 
can be directly used
in momentum space calculations (see Ref.~\cite{program}).

\begin{table}[h]
\begin{center}
\begin{math}
\begin{array}{l}
\hline \\
  \hat{\A}^R[1]  =  0 \\ \\
  \hat{\A}^R[\d_\mu]  =  0 \\ \\
  \hat{\B}^R[1]  =  \frac{1}{16\pi^2}  \log \frac{\bar{M}^2}{p^2-i\epsilon}
    \\ \\
  \hat{\B}^R[\d_\mu]  =  - \frac{1}{2} i p_\mu  
    \hat{\B}^R[1]
    \\ \\
  \hat{\B}^R[\Box]  =  0  \\ \\
  \hat{\B}^R[\d_\mu\d_\nu]  =   
    - \frac{1}{3} (p_\mu p_\nu-
    \frac{1}{4}p^2\delta_{\mu\nu}) \hat{\B}^R[1]
    - \frac{1}{288\pi^2} (p_\mu p_\nu-p^2\delta_{\mu\nu}) 
    \\ \\
\hline \\
\end{array} 
\end{math}
\end{center}
\caption{Fourier transforms of massless
one- and two-point renormalized basic functions. The small 
imaginary part 
in the logarithms here and in the next table, 
allows to analitically continue these formulae into 
the Minkowsky region ($p^2 < 0$).}
\end{table}

%
\begin{table}[h]
\begin{center}
\begin{math}
\begin{array}{l}
\hline \\
  \hat{\A}_m^R[1]  =   \frac{1}{16\pi^2} m^2 
    (1-\log \frac{\bar{M}^2}{m^2})  \\ \\
  \hat{\A}_m^R[\d_\mu]  =  0 \\ \\
  \hat{\B}^R_{m_1m_2}[1]  =  \frac{1}{16\pi^2} \left\{
    \frac{m_2^2-m_1^2}{p^2} \log \frac{m_2}{m_1}
    + \log \frac{\bar{M}^2}{m_1 m_2} + C_{m_1m_2}(p) \right\}
     \\ \\
  \hat{\B}_{m_1m_2}^R[\d_\mu]  =  -\frac{1}{2} i p_\mu
     \left[ \hat{\B}^R_{m_1m_2}[1] +
     \frac{1}{16\pi^2} \frac{1}{p^2} \left(
     m_1^2-m_2^2 +  \log \frac{m_1}{m_2} (m_1^2+m_2^2
     +\frac{(m_1^2-m_2^2)^2}{p^2}) \right. \right. \\
     \left. \left. \makebox[1cm]{} + 
     (m_1^2-m_2^2) C_{m_1m_2}(p)
     \right) \right]
     \\ \\
  \hat{\B}_{m_1m_2}^R[\Box]  = 
    m_2^2 \hat{\B}^R_{m_1m_2}[1] - \hat{\A}_{m_1}^R[1]
     \\ \\
  \hat{\B}_{m_1m_2}^R[\d_\mu\d_\nu]  =  
     -\frac{i}{2} \left( p_\mu \hat{\B}_{m_1m_2}^R[\d_\nu] -
     p_\nu \hat{\B}_{m_2m_1}^R[\d_\mu] \right)
     + \frac{1}{8} \delta_{\mu\nu} \left(
     \hat{\B}_{m_1m_2}^R[\Box] + \hat{\B}_{m_2m_1}^R[\Box] \right) \\
     \makebox[1cm]{} -
     \frac{1}{3} (p_\mu p_\nu - \frac{1}{4}\delta_{\mu\nu} p^2)
     \hat{\B}_{m_1m_2}^R[1] \\
     \makebox[1cm]{} -
     \frac{1}{6} \left\{ \left[
     m_1^2 \, \I^{(3)}_{m_2m_2m_1}[k_\mu k_\nu - 
     \frac{1}{4} \delta_{\mu\nu} k^2](0,p)
     + m_2^2 \, \I^{(3)}_{m_1m_1m_2}[k_\mu k_\nu -
     \frac{1}{4} \delta_{\mu\nu} k^2](0,p)
     \right] \right. \\
     \left. \makebox[1cm]{} - \left[ 
     m_1^2 \, \I^{(3)}_{m_2m_1m_1}[k_\mu k_\nu - 
     \frac{1}{4} \delta_{\mu\nu} k^2](p,p)
     + m_2^2 \, \I^{(3)}_{m_1m_2m_2}[k_\mu k_\nu - 
     \frac{1}{4} \delta_{\mu\nu} k^2](p,p)
     \right] \right\} \\
     \makebox[1cm]{} -
     \frac{1}{16\pi^2} \left[
     \frac{1}{18} (p_\mu p_\nu - \delta_{\mu\nu} p^2) -
     \frac{1}{8} (m_1^2+m_2^2) \delta_{\mu\nu} \right]
     \\ \\ \\
  \makebox[2cm][l]{where} \\ \\
    \makebox[1cm]{} C_{m_1m_2}(p)  =   
    - \frac{\lambda^{1/2}}{2 p^2}
    \left( \log (p^2+m_1^2+m_2^2+\lambda^{1/2}+i \epsilon) - 
    \log (p^2+m_1^2+m_2^2-\lambda^{1/2} - i \epsilon) \right)
     \, ,
    \\ \\
    \makebox[1cm]{} \lambda = \left( p^2+(m_1+m_2)^2 \right)
    \left( p^2+(m_1-m_2)^2 \right)
     \\  \\ 
\hline    
\end{array} 
\end{math}
\end{center}
\caption{Fourier transforms of massive
one- and two-point renormalized basic functions.
The integrals $\I^{(3)}$ are defined in the text.
The momentum $k$ is the integration variable.}
\end{table}
\begin{table}[h]
\begin{center}
\begin{math}
\begin{array}{l}
\hline \\
  \hat{\T}_{m_1m_2m_3}^R[\Box]  = 
    m_3^2 \, \I^{(3)}_{m_1m_2m_3}[1](p_x,p_y) - 
    \hat{\B}_{m_1m_2}^R[1](p_x-p_y) \\ \\
  \hat{\T}_{m_1m_2m_3}^R[\d_\mu\d_\nu]  =  
    - \I^{(3)}_{m_1m_2m_3}[k_\mu k_\nu-
    \frac{1}{4}\delta_{\mu\nu} k^2](p_x,p_y)
    + \frac{1}{4} \delta_{\mu\nu} 
    \hat{\T}_{m_1m_2m_3}^R[\Box](p_x,p_y)
    -\frac{1}{128\pi^2} \delta_{\mu\nu} \\ \\
  \hat{\T}_{m_1m_2m_3}^R[\Box\d_\mu]  =  
    - i m_3^2 \, \I^{(3)}_{m_1m_2m_3}[k_\mu](p_x,p_y) - 
    \hat{\B}^R_{m_1m_2}[\d_\mu](p_x-p_y) 
    + i p_{y\,\mu} \hat{\B}^R_{m_1m_2}[1](p_x-p_y)
    \\ \\  
  \hat{\T}_{m_1m_2m_3}^R[\d_\mu\d_\nu\d_\rho]  =  
    i \,\I^{(3)}_{m_1m_2m_3}[k_\mu k_\nu k_\rho-\frac{1}{6}
    (\delta_{\mu\nu} k_\rho + \delta_{\mu\rho} k_\nu
    +\delta_{\nu\rho} k_\mu) k^2](p_x,p_y) \\
    \makebox[1cm]{}
    + \frac{1}{6} \left(
    \delta_{\mu\nu} \hat{\T}_{m_1m_2m_3}^R
    [\Box\d_\rho](p_x,p_y) + \delta_{\mu\rho} 
    \hat{\T}_{m_1m_2m_3}^R[\Box\d_\nu](p_x,p_y) \right. \\
    \left. \makebox[1cm]{} +
    \delta_{\nu\rho} 
    \hat{\T}_{m_1m_2m_3}^R[\Box\d_\mu](p_x,p_y) \right) \\
    \makebox[1cm]{}
    +  \frac{i}{576\pi^2} \left(
    \delta_{\mu\nu} (p_{x\,\rho}+p_{y\,\rho}) +
    \delta_{\mu\rho} (p_{x\,\nu}+p_{y\,\nu}) +
    \delta_{\nu\rho} (p_{x\,\mu}+p_{y\,\mu}) \right)
    \\ \\
  \hat{\Q}_{m_1m_2m_3m_4}^R[\Box\Box]  = 
    - m_4^2 \, \I^{(4)}_{m_1m_2m_3m_4}[k^2](p_x,p_y,p_z) + 
    p_z^2 \, \I^{(3)}_{m_1m_2m_3}[1](p_x-p_z,p_y-p_z) \\
    \makebox[1cm]{}
    + 2  p_{z\,\rho} 
    \, \I^{(3)}_{m_1m_2m_3}[k_\rho](p_x-p_z,p_y-p_z) -
    \hat{\T}_{m_1m_2m_3}^R[\Box](p_x-p_z,p_y-p_z) \\ \\
  \hat{\Q}_{m_1m_2m_3m_4}^R[\Box\d_\mu\d_\nu]  = 
    - m_4^2 \, \I^{(4)}_{m_1m_2m_3m_4}[k_\mu k_\nu](p_x,p_y,p_z) 
    + p_{z\,\mu} p_{z\,\nu} 
    \, \I^{(3)}_{m_1m_2m_3}[1](p_x-p_z,p_y-p_z) \\
    \makebox[1cm]{}  +
    p_{z\,\mu} 
    \, \I^{(3)}_{m_1m_2m_3}[k_\nu](p_x-p_z,p_y-p_z) +
    p_{z\,\nu} 
    \, \I^{(3)}_{m_1m_2m_3}[k_\mu](p_x-p_z,p_y-p_z) \\
    \makebox[1cm]{} -
    \hat{\T}_{m_1m_2m_3}^R[\d_\mu\d_\nu](p_x-p_z,p_y-p_z) 
    \\ \\
  \hat{\Q}_{m_1m_2m_3m_4}^R[\d_\mu\d_\nu\d_\rho\d_\sigma]   =  
    \I^{(4)}_{m_1m_2m_3m_4}[k_\mu k_\nu k_\rho k_\sigma -
    \frac{1}{24}(\delta_{\mu\nu}\delta_{\rho\sigma} +
    \delta_{\mu\rho}\delta_{\nu\sigma}+
    \delta_{\mu\sigma}\delta_{\nu\rho}) k^4]
    (p_x,p_y,p_z) \\
    \makebox[1cm]{} +
    \frac{1}{24} (\delta_{\mu\nu}\delta_{\rho\sigma} +
    \delta_{\mu\rho}\delta_{\nu\sigma}+
    \delta_{\mu\sigma}\delta_{\nu\rho})
    \left(\hat{\Q}_{m_1m_2m_3m_4}^R[\Box\Box](p_x,p_y,p_z) + 
    \frac{5}{96\pi^2} \right) \\ \\
\hline
\end{array}
\end{math}
\end{center}
\caption{Fourier transforms of three- and four-point
renormalized basic functions.  
$p_x$, $p_y$ and $p_z$ are the conjugate
momenta of the coordinate variables $x$, $y$ and $z$.
The momentum-space basic functions on the \lhs\ are
assumed to depend on $p_x,p_y$ (for T functions)
and $p_x,p_y,p_z$ (for Q functions). The
integrals $\I^{(n)}$ are defined in the text.
The momentum $k$ is the integration variable.}
\end{table}

\end{document}